\documentclass[usenatbib,useAMS,twocolumn]{mn2e}
\usepackage{epsfig}
\usepackage{times}

\def\d{{\:\rm d}} 

\newcommand{\Teff}{\ensuremath{\mathit{T_{eff}}}}
\newcommand{\lgTeff}{\ensuremath{\mathit{\log{T_{eff}}}}}

\begin{document}

\title[Model atmospheres and thermal spectra of magnetized neutron stars]
{Model atmospheres and thermal spectra of magnetized neutron stars} 
\author[D. A. Lloyd]
{Don A. Lloyd \\ 
Harvard-Smithsonian
Center for Astrophysics, 60 Garden Street, Cambridge, MA 02138 \\
email: dlloyd@cfa.harvard.edu }

\maketitle
\begin{abstract}
Thermal surface radiation has been detected with X-ray instruments for
several neutron stars with high spectral, spatial and timing
resolution.  These observations allow for direct study of fundamental
properties of the source, but require model atmospheres and spectra
for careful interpretation.  We describe a robust and extensible
implementation of complete linearization for computing the spectra of
isolated cooling neutron stars for a broad range of temperature and
magnetic field.  Self-consistent spectra are derived for arbitrary
magnetic field geometries at $B\leq10^{14}\ \mathrm{G}$.
\end{abstract}

\section{Introduction}
\label{Sec:Intro}
Isolated neutron stars (NSs) are most commonly discovered as radio
pulsars and only infrequently as soft X-ray sources or through
associations with supernovae.  Cooling of the hot, isothermal NS core
$(T_{c}\sim 10^{9}\ \rmn{K})$ proceeds initially by neutrino
diffusion, followed by an extended period of thermal emission at soft
X-ray energies.  In this latter phase, conduction in the insulating
iron crust gives way to photon diffusion and free propagation of
radiation through the tenuous outer layers of the NS atmosphere.  The
thermal evolution of NS has been the subject of much theoretical
consideration since the seminal work of \citet{TsurutaPhD}, and
\citet{Chiu}.  In principal, measurement and interpretation of thermal
X-rays emitted at the NS surface provides direct evidence of the
star's surface composition, magnetic field properties, and, in the
absence of reheating, may constrain the cooling history of the NS
population or the equation of state of the stellar core.  In practice,
detection of these thermal X-rays is difficult.  The stellar
environment is energetic, particularly for young NS, and non-thermal
emission processes which originate in the active pulsar magnetosphere
dominate the surface radiation.  Young pulsars may be obscured by
diffuse non-thermal X-ray emission from the supernova remnant or from
bright compact nebulae (plerions) driven by pulsar winds.  Some NS are
exceptionally luminous in thermal X-rays possibly owing to either a
highly transparent surface composition or to reheating processes which
maintain their core temperature, while the thermal and (brighter)
non-thermal emissions for some cooling NS decline presumably at
different rates until the surface radiation can be distinguished at
late times.

The fundamental properties of NS thermal radiation are regulated by
the source luminosity, the strong surface gravity of the star, and the
composition of the atmosphere which is likely determined by either
accretion (from the interstellar medium, a fossil debris disk or
stellar companion) or fall-back of supernova ejecta.  Provided
accretion is neither too great nor continuous, gravitational settling
of heavy elements occurs on short timescales \citep*{BBR98}, leaving
the lightest elements to form the X-ray photosphere; only a tiny
fraction of the stellar mass in these light elements is necessary to
form an optically thick layer.  Spectral reprocessing by the light
element plasma atmosphere produces surface emission which differs
substantially from the blackbody function regardless of magnetic field
\citep*{Romani87,RR96,Zavlin96,Lloyd03}, and these discrepancies
motivate the accurate computation of model atmospheres and
spectra.  Tabulations of unmagnetized NS X-ray spectra are now
available in the XSPEC software \citep{Pavlov92,Zavlin96,Gaensicke02},
allowing for immediate comparison with the blackbody model for
$B\la10^{10}\ \mathrm{G}$.  The optical and near UV thermal flux shows
strong dependence on both composition \citep{Pavlov96b} and magnetic
field strength \citep{Lloyd03}.  The most persuasive models for NS
thermal radiation must reconcile the X-ray spectrum with optical and
UV fluxes whenever possible.

Several codes have been developed to calculate the
atmospheric structure and associated thermal spectrum for a variety of
NS surface compositions and magnetic field regimes.  The atmosphere is
described by the self-consistent solution to a system of radiative
transfer equations and equilibrium constraints.  Essentially all
methods for calculating radiative transport in magnetized NS
atmospheres proceed from the work of \citet{Gnedin}, who established
the connection between the density-matrix formalism and the (more
immediately tractable) formulation of radiative transport in two
coupled ``normal'' polarization modes; i.e., the limit of large
Faraday depolarization.  The plasma is strongly polarizing to X-ray
radiation at magnetic field strengths comparable to those of radio
pulsars.  The optical and UV spectra of such NS are also polarized for
modest $B$, but measurement at these wavelengths is difficult owing to
the presence of non-thermal radiation, absorption in the interstellar
medium, and the overall faintness of the sources.  The large plasma
anisotropies induced by magnetic fields strongly affect the angular
distribution of the emergent radiation field of the atmosphere, and a
complete model of the thermal X-ray emission of a typical NS requires
integration over intensity profiles from surface elements having local
temperature and magnetic field which vary across the stellar surface
\citep{Zavlin95}.  All modern strategies appeal to the full angle and
energy dependence of the plasma opacity and coupled radiation field.

Early efforts to model the X-ray emission from NS atmospheric plasma
were preceded by the development of the radiative transfer by the
normal mode approximation in uniform (isothermal) plasma slabs
\citep{KPS83} using the normal mode strategy of \citet{Gnedin} to
describe the radiative diffusion.  \citet{Shib92} applied the
diffusion method of \citet{KPS83} to NS atmospheres with fully ionized
H/He mixtures by implementing the constraints of hydrostatic and
radiative equilibrium.  \citet{Pavlov94} computed spectra for
generally oblique magnetic fields the NS atmosphere from the thermal
structures from \citet{Shib92} by recalculating the plasma opacities
and making a formal integration of the transfer equation from the
prior solution.  \citet{Pavlov95} improved the diffusion results by
using them as an approximate solution to the temperature correction
procedure of \citet{AuerMihalas68}; this work also included a
prescription for the ionization equilibrium of the Hydrogen plasma
(see also \citet{ZP02}).

\citet{Romani87} described the first self-consistent NS model
atmospheres for weak magnetic fields using the Los Alamos Opacity
Library and some simplifying assumptions about the relative
contribution of thermal and scattering emissivities.  In this method,
the radiation field is derived from application of the Milne integral
operator to the source function and temperature corrections to the
atmospheric structure were evaluated by the Lucy-Uns\"old method
\citep{Mihalas78}.  \citet{RR96} revisited this work using the OPAL
tabulations \citep{OPAL} while \citet{RRM97}, using essentially the
same formalism, extended their opacity model to roughly approximate
the bound state transitions in magnetized iron, and included electron
scattering and free-free absorption in the normal mode approximation.
Each of these efforts revealed the extent of compositional effects on
the thermal X-ray spectra of He, Fe and cosmic abundance atmospheres.
More recently, \citet{Gaensicke02} have revisited the low-field regime
using these methods to consider reheating in the outer atmospheric
layers of some NS.  \citet{Werner00} have taken a different
computational approach, deriving the first non-LTE model calculations
in the weak field regime using the Accelerated Lambda Iteration
combined with the Opacity Project cross sections; these authors find
modest corrections due to NLTE effects in light element plasma at low
temperatures, but these are potentially more significant for magnetic
fields stronger than they consider.  Some recent modeling efforts have
specialized to the calculation of light element ``magnetar'' spectra
$(B\ga 10^{14}\ \mathrm{G})$ which include some prescription for
vacuum polarization effects which are particularly relevant for $B\ga
4.4\times 10^{13}\ \rmn{G}$.  \citet{Zane00} considered in particular
atmospheres heated externally at low accretion rates, generalizing
energy balance in the atmosphere to account for energy deposition by
charged particles.  \citet{Ozel01} and \citet{HoLai01} have each
generalized the the Lucy-Uns\"old method to the normal mode
approximation to derive model spectra for highly magnetized, light
element NS plasma.  \citet{HoLai01} also considered the diffusion
equation solutions to perpendicular (normal) and parallel magnetic
field geometries in the intense field regime.

This article describes an implementation of the classical technique of
complete linearization (CL) to the modeling of elementary NS
atmospheres in radiative equilibrium.  We restrict our derivation to
model atmospheres with pure Hydrogen composition in the limit of
complete ionization, but the formulation is immediately useful to
admixtures of Helium.  The method is robust, flexible with respect to
the range of model parameters for which it can be applied, and is
extensible to detailed statistical equilibrium for partially ionized
models and complete Stokes transport.  Self-consistent model results
for magnetic fields $B\leq10^{14}\ \rmn{G}$ and of arbitrary field
orientation, and a range of flux temperatures characteristic of NS
observed in thermal X-rays have been calculated with this methodology.
The article is organized as follows: The magnetized plasma opacity is
derived in \S\ref{Sec:Opacity}, with attention to the role of protons
and vacuum effects in radiative processes.  The radiative transfer
equations and their constraints are derived in \S\ref{Sec:RTE}.  The
grid methods and solution techniques are described in
\S\ref{Sec:Solution}.  An abbreviated selection of model results are
presented in \S\ref{Sec:Applications}, and emphasize the angular
intensity properties and spectral characteristics, including
polarization for the weak, intermediate and strong field regimes.
Extensions and limitations of the present method are outlined in
\S\ref{Sec:Discussion}.

\section{Opacity and emissivity of magnetized plasma}
\label{Sec:Opacity}
The detailed polarizing properties of a light element magnetized
plasma can be derived from its dielectric response to electromagnetic
waves. This presentation is similar to that found in \citet{Vent79}
and \citet{MeszVent79}.  The equation for an electromagnetic wave
propagating in a medium of refractive index $N$ along the direction
$\hat{k}$, including the vacuum permeability, is
\begin{equation}
\label{eq:waveeqn}
\left[-a N^{2}(\vec{\vec{1}}-\hat{k}\hat{k})+h
N^{2}(\hat{k}\times\hat{b})(\hat{k}\times\hat{b})+\epsilon \right]
\vec{E} = 0
\end{equation}
where $\hat{b}=(\sin{\theta},0,\cos{\theta})$ is the unit vector in
the direction of $\vec{B}$, and $a$ and $h$ are defined below.  For
uniformly magnetized cold plasma, the vacuum corrected dielectric in
the $\vec{B}\parallel\hat{z}$ frame is
\begin{equation}
\label{eq:vacplasdielect}
\epsilon' = \left( \matrix{ \varepsilon & i \,g & 0 \cr -i \,g &
\varepsilon & 0 \cr 0 & 0 & \eta \cr } \right)
\end{equation}
and has tensor elements
\begin{eqnarray}
\varepsilon &=& a-\sum_{s}\frac{\lambda_{s}
v_{s}}{\lambda_{s}^{2}-u_{s}} \\ g &=&
-\sum_{s}\frac{v_{s}u_{s}^{1/2}}{\lambda_{s}^{2}-u_{s}} \\ \eta &=&
a+q-\sum_{s}v_{s}/\lambda_{s}
\end{eqnarray}
in which the plasma and cyclotron frequencies for each species enter
through the parameters
\begin{eqnarray}
v_{s} &=& (\omega_{p,s}/\omega)^2 \\ u_{s}^{1/2} &=&
(\omega_{c,s}/\omega)
\end{eqnarray}
for $\omega_{p,s}^{2}=4\upi e^{2} n_{s}/m_{s}$ and
$\omega_{c,s}=q_{s}B/(m_{s}c)$; note that $u_{s}^{1/2}$ is a signed
quantity.  The damping term $\lambda_{s} = 1 + i \nu_{s}/\omega$
splits the plasma dielectric into hermitian and anti-hermitian
contributions, but when evaluating the polarization of propagating
waves it is acceptable to consider only the hermitian part of the
dielectric.  The parameters $a,h$ and $q$ are functions solely of the
magnetic field strength:
\begin{eqnarray}
\label{eq:vacafunc}
a(x) &=& 1 + \frac{\alpha}{2\upi}\left[ -2 X_{0}(x) + x X_{0}'(x)
\right] \\ h(x) &=& \frac{\alpha}{2\upi}\left[ x^{2} X_{0}''(x) - x
X_{0}'(x) \right] \\
\label{eq:vacqfunc}
q(x) &=& -\frac{\alpha}{2\upi} X_{1}(x)
\end{eqnarray}
where $x\equiv{B_Q/B}$, $B_{Q}\equiv m_{e}^{2} c^{3}/(e \hbar)
\simeq4.4\times10^{13}$ G is the quantum critical field strength, and
the functions $X_{0}$ and $X_{1}$ are found in \citet{HH97}; the
expressions (\ref{eq:vacafunc}-\ref{eq:vacqfunc}) are valid for all
field regimes and revert to the results of \citet{MeszVent79} in the
weak field limit, $x>1$.  The tensor $\epsilon$ in the
$\hat{k}\parallel \hat{z}$ frame is recovered from
(\ref{eq:vacplasdielect}) by an orthogonal transformation of angle
$\cos{\theta}=\hat{k}\cdot\hat{b}$ about the $\hat{k}\times\hat{b}$
axis.

The polarization vector for an EM wave in mode $j$ in rotating
coordinates about the magnetic field is
\begin{equation}
\label{eq:polzvec}
\vec{e}^{j} = \left(
\frac{1}{\sqrt{2}}\left[K_{x}^{j}\cos{\theta}-K_{z}^{j}\sin{\theta}\pm
i\right], K_{z}^{j}\cos{\theta}+K_{x}^{j}\sin{\theta} \right)
\end{equation}
where the component indices are $(\pm,0)$, respectively.  The normal
modes are usually referred to as ordinary (O) and extraordinary (X);
waves propagating in the ordinary mode are largely unaffected by the
magnetic field, while X mode radiation with $\omega<\omega_{c,e}$
propagates with a reduced refractive index.  Adopting the notation
$j=(1,2)$ to label mode (X,O) respectively, the transverse ellipticity
may be expressed compactly:
\begin{equation}
\label{eq:Kx_short}
K_{x}^{j} = \frac{i \zeta}{b+(-1)^{3-j} \rmn{sgn}(y)
\sqrt{b^{2}+\zeta}}
\end{equation}
where
\begin{eqnarray}
\label{eq:bparam}
b &=& \frac{x \sin^{2}{\theta}}{2 g \eta \cos{\theta}} \\
\label{eq:xparam}
x &=& \varepsilon^{2}- g^{2} -\varepsilon \eta (1-h/a) \\ y &=&
\frac{x\sin^{2}{\theta}}{a\left(\varepsilon\sin^{2}{\theta}+
\eta\cos^{2}{\theta}\right)} \\ \zeta &=& 1-h \sin^{2}{\theta}/a
\end{eqnarray}
and the longitudinal polarization component has ellipticity given by:
\begin{equation}
\label{eq:Kz}
K^{j}_{z} = -\frac{\left[i
g+(\varepsilon-\eta)K_{x}^{j}\cos{\theta}\right]\sin{\theta}}{\varepsilon
\sin^{2}{\theta} +\eta\cos^{2}{\theta}}
\end{equation}

\subsection{Cross sections}
\label{Sec:Cross}
The principal opacity sources in a fully ionized, light element plasma
are the free-free absorption and ordinary Thomson scattering.  The
differential cross-section for scattering by electrons of photons with
momentum $k$ in polarization mode $i$ into momentum $k'$ and mode $j$
is \citep{Vent79}:
\begin{equation}
\label{eq:rawdxs}
\left(\frac{d\sigma_{ij}^{kk'}}{d\Omega}\right) =
\frac{k'}{k}r_{0}^{2}\left|\hat{e}',\Pi^{(p)} \cdot\hat{e} \right|^{2}
\end{equation}
where $r_{0}=e^{2}/(mc^{2})$ is the classical electron radius (cm).
Superscript $(p)$ denotes plasma components only; e.g.
the plasma polarization tensor is
\begin{equation}
\label{eq:polztensor}
\Pi^{(p)} = v_{e}^{-1}\left(\delta-\varepsilon^{(p)}\right)
\end{equation}
The eigenvalues of (\ref{eq:polztensor}) for the plasma component of
the total dielectric (\ref{eq:vacplasdielect}) are
\begin{eqnarray*}
\label{eq:polzeigen}
\pi_{\pm} &=& v_{e}^{-1}\left[ 1-\epsilon^{(p)}\pm g^{(p)} \right] \\
\pi_{0} &=& v_{e}^{-1} \left[ 1-\eta^{(p)}\right]
\end{eqnarray*}
To excellent approximation, the squared eigenvalues can be written
compactly as a rational expression in terms of the plasma
constituents; including the damping term
$\gamma_{s}\equiv\nu_{s}/\omega$ these are:
\begin{equation}
\label{eq:polzeigenrational}
\pi_{p}^{2} = \frac{\left(\sum_{s} m_{e}/m_{s}
\right)^{2}}{\prod_{s}\left[\left(1- p u_{s}^{1/2}\right)^{2}+
\gamma_{s}^{2} \right]}
\end{equation}
For coherent scattering, eqn (\ref{eq:rawdxs}) is a sum over
polarization components $p$, and has a scale fixed by the Thomson
cross section $\sigma_{T}=8\upi r_{0}^{2}/3$:
\begin{equation}
\label{eq:diffxs}
\left(\frac{d\sigma_{ij}^{kk'}}{d\Omega}\right) = \sigma_{T}
\left[\frac{3}{8\upi} \sum_{p=-1}^{+1} \left|e_{p}^{j}(k')\right|^{2}
\left|e_{p}^{i}(k)\right|^{2} \pi_{p}^{2}\right]
\end{equation}
Integration of (\ref{eq:diffxs}) over final momenta $k'$ yields the
partial cross-section to mode $i$ photons which scatter into momentum
$k$ and mode $j$
\begin{equation}
\label{eq:partialxs}
\sigma_{ij}(k) = \sigma_{T} \left[\frac{3}{8\upi} \int_{\Omega'}
\d\Omega' \sum_{p=-1}^{+1} \left|e_{p}^{j}(k')\right|^{2}
\left|e_{p}^{i}(k)\right|^{2} \pi_{p}^{2} \right]
\end{equation}
The bracketed term is the geometric correction to the scattering
process induced by the magnetic field in the mode conserving $(i=j)$
or mode crossing channels $(i\neq j)$.  The damping factor
$\gamma_{s}$ is obtained from the sum of the radiative and collisional
damping frequencies:
\begin{eqnarray*}
\nu_{s,rad} &=& 2(e\omega)^{2}/(3m_{s} c^{3}) \\ \nu_{s,coll} &=&
\alpha_{0}\nu_{rad}/(n_{s}\sigma_{T})
\end{eqnarray*}
where the free-free absorption coefficient $\alpha_{0}$ is defined
below.  Damping is only relevant near the resonant frequencies of
$\pi^{2}_{\pm}$, i.e., $\omega\simeq \omega_{c,s}$.

The total electron scattering cross-section to mode $i$ photons is
obtained by summing (\ref{eq:partialxs}) over final polarization $j$.
\begin{eqnarray}
\nonumber \sigma_{i}(k) &=& \sigma_{T} \left[\frac{3}{8\upi}
\int_{\Omega'} \d\Omega' \sum_{j=1}^{2} \sum_{p=-1}^{+1}
\left|e_{p}^{j}(k')\right|^{2} \left|e_{p}^{i}(k)\right|^{2}
\pi_{p}^{2} \right] \\
\label{eq:totalxs_2sums}
&\equiv& \sigma_{T} G_{i}(k)
\end{eqnarray}
The magnetic free-free absorption coefficient is similarly derived;
the total absorption for mode $i$ photons is:
\begin{eqnarray}
\nonumber \alpha_{i}^{\mathit{ff}}(k) &=& \alpha_{0}
\left[\frac{3}{8\upi} \int_{\Omega'} \d\Omega' \sum_{j=1}^{2}
\sum_{p=-1}^{+1}g_{p} \left|e_{p}^{j}(k')\right|^{2}
\left|e_{p}^{i}(k)\right|^{2} \pi_{p}^{2} \right] \\
\label{eq:totalff_2sums}
&\equiv& \alpha_{0} G_{i}^{\mathit{ff}}(k)
\end{eqnarray}
where $g$ is the Gaunt factor and $\alpha_{0}$ is the
absorption coefficient in the zero field limit, corrected for
stimulated emission:
\begin{equation}
\alpha_{0} = n_{e} n_{i} Z^{2}
\frac{4e^{6}}{3m_{e}hc}\left(\frac{2\upi}{3m_{e}kT}\right)^{1/2}
\nu^{-3} (1-e^{-h\nu/kT})
\end{equation}
The corresponding emissivity is related to the absorption coefficient
through the Einstein rate coefficients; i.e., Kirchoff's law:
\begin{equation}
\label{eq:bremms}
j^{i}_{\mathit{\nu,ff}} = n_{e} n_{i} Z^{2}
\frac{8e^{6}}{3m_{e}c^{3}}\left(\frac{2\upi}{3m_{e}kT}\right)^{1/2}
e^{-h\nu/kT} G_{i}^{\mathit{ff}}(k)
\end{equation}
Equation (\ref{eq:bremms}) is the correct thermal emissivity of
unmagnetized plasma; for arbitrarily magnetized plasma, the expression
must be multiplied by $0.5$ to obtain the emissivity \textit{per
mode}.  In the polarizing medium, thermal emissivity is greater for
the more opaque mode.  It is convenient to scale the opacity and
emissivity by the local mass density:
\begin{equation}
\chi_{\nu}^{i} = \frac{\alpha_{i}^{\mathit{ff}} + n_{e}
\sigma_{i}}{\rho}\ \ \rmn{cm^{2}\ g^{-1}}
\end{equation}
\begin{equation}
\eta_{\nu}^{i} = \frac{j_{\nu,th}^{i}+j_{\nu,sc}^{i}}{\rho}\ \ 
\rmn{erg\ g^{-1}\ s^{-1}\ Hz^{-1}\ sr^{-1}}
\end{equation}
The thermal emissivity $j^{th}_{i}$ is given by eqn (\ref{eq:bremms}),
and the scattering emissivity in mode $i$ is
\begin{equation}
j^{i}_{\nu,sc}(k) = n_{e} \int_{\Omega'} \d\Omega' \sum_{j=1}^{2}
u^{j}(k') \left(\frac{d\sigma_{ji}^{k'k}}{d\Omega'}\right)
\end{equation}
where the integration is performed over incoming photon momenta.

For unmagnetized plasma, the Gaunt factors and their derivatives are
evaluated from the fitting formulae of \citet{Itoh00} or the
tabulations of \citet{Sutherland98}.  The magnetic Gaunt factors and
their temperature derivatives are evaluated by the integral
expressions found in \citet{KirkMesz80}; this calculation proceeds
from an average of the ground state Coulomb matrix elements over a
one-dimensional electron distribution in the Landau ground level,
which adequately describes the available electronic states in highly
magnetized plasma.  In weak magnetic fields, we expect that the
anisotropy of the electron distribution is sufficiently mild in this
regard to allow substitution of the Gaunt factors $g_{\pm,0}$ with the
zero-field values.  The transition from the classical to quantizing
magnetic field regimes is characterized by the parameter
\begin{equation}
\label{eq:landaupops}
\ell=k\Teff/(\hbar \omega_{c,e}) \simeq 7.4\times10^{3} \Teff/B
\end{equation}
which is computed from the effective temperature $\Teff$ defined in
\S\ref{Sec:RTE} and neglecting variation in local temperature in a
realistic atmosphere.  In this formulation the Gaunt factors are
incompletely calculated over a range of $(\Teff,B)$ corresponding to
the transition between the classical field and strongly magnetized
regimes $\ell\sim\mathcal{O}(1)$.

The geometric correction factors $G_{i}(k)$ and
$G_{i}^{\mathit{ff}}(k)$ in (\ref{eq:totalxs_2sums}) and
(\ref{eq:totalff_2sums}) are fundamental quantities in the opacity
calculations which fully describe all resonant behaviors of the
magnetized plasma through self-consistent inclusion of ions and vacuum
in the dielectric (\ref{eq:vacplasdielect}), and which are computed
explicitly for all model photon energies and trajectories.  The cross
sections are invariant to reversal of the magnetic field direction,
and each cyclotron line has an effective width $\Delta\omega_{c,s}
\propto \omega_{c,s}$ in excess of either the thermal or natural width
$\gamma_{s}$ of the resonance.  Note that our choice for $\Pi^{(p)}$
generalizes the polarization tensor for the multi-component plasma and
produces the expected resonances at the electron and ion cyclotron
frequencies.  The vacuum dielectric influences only the calculation of
the polarization vectors (\ref{eq:polzvec}) with consequences
discussed below.  Cross sections for direct absorption and scattering
by ions are reduced from their respective electronic values by a
factor $\mathcal{O}(m_{e}/m_{p})^{2}$ and may be neglected without
consequence when calculating the opacity.  The opacity to radiation in
parallel propagation to the magnetic field is approximately identical
in each mode, regardless of the magnitude of the vacuum corrections to
the plasma response; substantial differences in the differential cross
sections can produce unequal intensities and finite polarization along
magnetic field lines.

Electron and ion cyclotron absorption each result from different
functional properties of the magneto-geometric factors.  Both features
are sensitive to projection of the X mode polarization vectors onto
the circular basis about the magnetic field (\ref{eq:polzvec}).  In
the electron feature, projection over all angles onto the resonant
eigenvalue $\pi_{-}^{2}$ is complete; therefore only radiation in X
mode is resonantly absorbed while the O mode has no projection on
$\pi_{-}^{2}$ and is consequently insensitive to resonant absorption.
At frequencies $\omega\ll\omega_{c,e}$ the $\hat{e}^{X}_{0}$ is
everywhere zero except near either ion cyclotron frequency where it
resonantly acquires a finite projection.  Owing to the large relative
opacity to waves polarized parallel to the magnetic field at these
frequencies, it is this ``polarization vector'' resonance which
dominates the total opacity to X mode radiation in the ion cyclotron
resonance. The polarization structure in the resonance is complicated
further by the abrupt exchange of $\pm$ components at a critical
frequency near $\omega_{c,p}$ \citep{Bulik96}, but the magnitude of
the circular components remains symmetric across the feature.

The magnetic field provides a preferred orientation to scattering and
absorption processes, resulting in anisotropic plasma response and
finite polarization of propagating radiation.  The magnitude of these
effects scale principally with $u_{e}=(\omega_{c,e}/\omega)^{2}$.  Ion
cyclotron features arise independent of vacuum corrections from a
parametric resonance between electrons and ions in the dielectric
tensor.  Moreover, any additive contributions to the total plasma
dielectric (e.g. line or ionization transitions) introduce additional
polarization resonances.  The electronic and ionic plasma constituents
cannot be consistently treated as independent entities.

\subsection{Vacuum resonance and mode ambiguity}
\label{Sec:VP}
In the previous section, formation of the ion cyclotron resonance was
shown to arise primarily from a critical parametric phenomenon in the
spectrum of the polarization vector components, $\hat{e}_{p}^{j}$.
This is one example of the more generic competition between multiple
sources of polarizability in magnetized plasma.  More generally, any
additive component to the dielectric (\ref{eq:vacplasdielect})
(e.g. transitions in neutral atoms) can produce other resonant
behaviors, and introduce additional complexity to the spectrum of
critical behavior.  In intense magnetic fields $(B\ga B_{Q})$, vacuum
polarizability has a significant influence on the X-ray plasma
opacities, and possesses parametric polarization resonances analogous
to those of ion cyclotron lines.  Birefringence of the magnetized
vacuum has been investigated by many authors (e.g. \citet{Adler71} for
review and \citet{Pavlov79} for discussion relevant to NS
applications); strictly, the effect remains finite for arbitrarily
weak fields although, to excellent approximation, the vacuum
corrections to weakly magnetized plasma are identical to the
uncorrected (pure) plasma.  Our discussion of the vacuum effect
proceeds from the work of \citet{Soffel83}.

Inspection of (\ref{eq:Kx_short}) reveals that the transverse
ellipticities of the two modes are identical for arbitrary $\theta$
when $b=\pm i\sqrt{\zeta} \rightarrow K_{x}^{j}=\pm b$. This is
distinct from (e.g.) the case of parallel propagation
$(\hat{k}\parallel \vec{B})$ in which the modes are circular, having
identical \textit{magnitude}, but opposite and unambiguous handedness.
Where this occurs in a radiative transfer calculation, it may be
impossible to uniquely specify the polarization content of the
affected radiation field in two normal modes, sometimes called ``mode
collapse''.  To understand the limitation of our model calculations,
it is necessary to investigate the plasma conditions under which the
mode definition (\ref{eq:Kx_short}) is rendered ambiguous.  To deduce
the critical density at which mode collapse occurs, it is sufficient
to study the necessary condition that $\rmn{Re}(b)=0$ for arbitrary
propagation angles, i.e. where $x=0$ (eqns
(\ref{eq:bparam}-\ref{eq:xparam})).  To first order in the
fine-structure constant, this condition is satisfied to good
approximation when
\begin{equation}
\label{eq:crit_vac_plas_density}
v_{res}=(q-h)\left(\frac{u_{e}-1}{u_{e}}\right)
\end{equation}
The critical density (\ref{eq:crit_vac_plas_density}) is undefined for
$\omega>\omega_{c,e}$ and photon frequencies in this range are
insensitive to vacuum resonance effects at all plasma densities.  For
photons $\omega<\omega_{c,e}$, the resonant density
(\ref{eq:crit_vac_plas_density}) delineates a transition from the
plasma dominated dielectric response into the vacuum dominated regime.
For polarization vector calculations which invoke only real valued
$b$, the modes acquire a finite but insubstantial measure of
non-orthogonality when $b=0$:
\[
\left| \hat{e}^{1^{*}}_{t}\cdot \hat{e}^{2}_{t}\right| = h
\sin^{2}{\theta}/a \ll 1
\]
where the transverse components $\hat{e}_{t}^{j}
\equiv(\vec{\vec{1}}-\hat{k}\hat{k})\cdot\hat{e}^{j}$.  When computing
radiative transfer in such a medium, the mode identities remain
distinct.  Equation (\ref{eq:crit_vac_plas_density}) is a quadratic
expression for two critical frequencies which are responsive to the
vacuum resonance at a given plasma density:
\begin{equation}
\label{eq:crit_vac_freqs}
\omega^{2}_{\pm} = \frac{\omega_{c,e}^{2}}{2} \left[1 \pm
\sqrt{1-\frac{4}{q-h}\left(\frac{\omega_{p,e}}{\omega_{c,e}}\right)^{2}}
\right]
\end{equation}
$(q>h)$ for all $B$ and the critical frequencies are defined only for
$\omega_{p,e}^{2} < \omega_{c,e}^{2}(q-h)/4$.  For $v>v_{res}$ the
opacity is essentially unchanged from the pure plasma calculation and
this regime is said to be \textit{plasma dominated}.  In the
\textit{vacuum dominated} regime $(v<v_{res})$ the modes become
strictly linear, albeit slightly non-orthogonal, for all propagation
angles $\theta>0$, while both modes are fully circular for $\theta=0$
in all circumstances.  An additional consequence of conversion to
linear polarization is that the angular dependence to the opacity in
each mode is quenched in the vacuum regime.  This implies that
radiation decoupled from the plasma in the regime $v_{res}\ga
v(\tau_{\nu}=1)$ will also become isotropized but, as we find in
\S\ref{Sec:Apps:Magnetar}, this possibility is not realized in model
atmosphere solutions.

The imaginary part of $b$ is approximately
\begin{equation}
\label{eq:imb_general}
\rmn{Im}(b) \simeq \gamma_{e}\sqrt{u}\left[ 1-\frac{2}{u-1}\right]
\frac{\sin^{2}{\theta}}{2 \cos{\theta}}
\end{equation}
For magnetic field strengths capable of inducing the resonant
phenomenon in NS atmospheres, the critical frequency
$\omega_{+}\simeq\omega_{c,e}$ is well above the limit of measurable
thermal emission.  Evaluating (\ref{eq:imb_general}) at the low
frequency resonance and taking $\rmn{Im}(b)=\sqrt{\zeta}$, we find the
critical angle at which the modes collapse at $\omega_{-}$ is
$\cos{\theta_{-}} \simeq 0.5\,\rmn{Im}(b)\ll 1$, or $\theta_{-} \simeq
\upi/2$.$\rmn{Im}(b)$ is a rapidly varying function near the critical
angle, and for only a very narrow range of $\theta\simeq\theta_{-}$
will the polarization modes become indistinguishable.

In the vacuum regime, the two critical frequencies are $\omega_{+}
\simeq \omega_{c,e}$ and $\omega_{-} < \omega_{+}$.  \textit{Both}
polarization modes become sensitive to the electron cyclotron
resonance in the vacuum dominated regime; this is however of no
special consequence for our NS spectra where $\omega_{c,e}$ is
unobservable for magnetic fields capable of producing the resonance
behavior.  From eqns (\ref{eq:crit_vac_plas_density}) and
(\ref{eq:crit_vac_freqs}) we find that, for a particular magnetic
field strength $B$, there are two critical frequencies, while, for any
particular photon energy, the vacuum resonance core is found at a
specific plasma density $n_{e}$.

Presently, no NS model atmosphere calculations fully specify the
mixing of polarization at the mode collapse points
$(\omega_{\pm},\theta_{\pm})$.  \citet{HoLai02} have pursued
calculations invoking the two extreme limits of ``no collapse'' and of
wholesale exchange of polarization identities through all photon
trajectories, with the correct solution presumably intermediate to
these extremes; these authors have since drawn conclusions similar to
those above which favor the limit of distinct polarization modes
\citep{HoLai03}.  From the preceding discussion, it is acceptable to
assume that mode-collapse will have at best a modest effect on the
emergent NS spectrum regardless of the magnitude of the effect within
the $\theta_{-}$ annulus, and we consider only the non-collapse
limit in the remainder of this article.  The mode ambiguity is related
to the breakdown of the Faraday depolarizing limit, and a more
exacting solution will likely require radiative transport of the full
Stokes vector, and careful treatment of the anti-hermitian components
of the plasma dielectric.

\section{Radiative transfer and neutron star atmospheres}
\label{Sec:RTE}
The fundamental parameters of the model atmosphere are the surface
gravity $g_{s}$, the total flux propagating through the atmosphere
$\mathcal{F}_{cons}$, the magnetic field strength and orientation
$\vec{B}$, and the plasma composition.  For a given equation of state,
these parameters uniquely specify the model.

The surface gravity of a relativistic star is
\[
g_{s} \equiv \gamma \frac{GM}{R^{2}} = \frac{GM/R^{2}}{\sqrt{1-2GM/R}}
\]
and for $M,R$ characteristic of NSs, $g_{s} \sim \mathcal{O}(10^{14})\
\rmn{cm~s^{-2}}$.  The large surface gravity gives an atmospheric
scale height $d\ll R$, and gradients in the gravitational and magnetic
fields of the star may be neglected for the purpose of calculating the
atmospheric structure.  The thin plasma atmosphere is well described
by a semi-infinite plane parallel geometry.  We do not consider
extended atmospheres or those with bulk motions.

In the absence of heat sources or sinks, the total energy flux is a
conserved property, and we define the effective temperature
$T_{\mathit{eff}}$ in terms of this conserved flux:
\[
\sigma_{SB} T_{\mathit{eff}}^{4} = \mathcal{F}_{\mathit{cons}}
\]
for $\sigma_{SB}$ the Stefan-Boltzmann constant.  Conductive transport
at densities characteristic of NS atmospheres $(\rho \sim
0.1-10\:\rmn{g\:cm^{-3}})$ is inefficient compared to radiative
diffusion, while convective instabilities, already suppressed in the
limit of complete ionization, are further inhibited by Ohmic
dissipation in strong magnetic fields.  Consequently, neither
conductive nor convective transport is included in our calculations.

In a semi-infinite, plane parallel geometry, the steady-state
transport of radiation in mode $j$ for photon momentum \textbf{k} at
any particular depth in the stellar medium is described by the
differential equation
\begin{equation}
\label{simple_RTE}
\mu\frac{dI^{j}_{\nu}(\bmath{k})}{dm} = \chi^{j}_{\nu}({\bmath{k}})
I^{j}_{\nu}({\bmath{k}}) - \eta^{j}_{\nu}({\bmath{k}})
\end{equation}
The momentum vector \textbf{k} is assumed to be outward pointing, and
is specified by the surface latitude $\mu=\cos{\theta_{k}}$ and
azimuthal angle $\phi$; the latter is measured with respect to a
reference direction to be defined later.  All functional dependence on
the magnetic field vector is subsumed in the formulation of
$\chi^{j}_{\nu}({\bmath{k}})$ and $\eta^{j}_{\nu}({\bmath{k}})$.  It
is advantageous to write the radiative transfer equation
(\ref{simple_RTE}) under transformation to the pair of conjugate
intensity (Feautrier) variables:
\begin{eqnarray}
\label{eq:Feaut_var_u}
u^{j}_{\nu}({\bmath{k}}) &=& \frac{1}{2} \left[
I^{j}_{\nu}({\bmath{k}}) + I^{j}_{\nu}({\bmath{-k}}) \right] \\
\label{eq:Feaut_var_v}
v^{j}_{\nu}({\bmath{k}}) &=& \frac{1}{2} \left[
I^{j}_{\nu}({\bmath{k}}) - I^{j}_{\nu}({\bmath{-k}}) \right]
\end{eqnarray}
where for ${\bmath{k}}= {\bmath{k}}(\mu,\phi)$, ${\bmath{-k}}=
{\bmath{k}}(-\mu,\upi-\phi)$ and $\mu>0$.  The transfer equations in
these variables are
\begin{eqnarray}
\label{eq:Feaut_eqn_v}
\mu \frac{dv^{j}_{\nu}({\bmath{k}})}{dm} &=&
\chi^{j}_{\nu}({\bmath{k}}) u^{j}_{\nu}({\bmath{k}})-
\eta^{j}_{\nu}({\bmath{k}}) \\
\label{eq:Feaut_eqn_u}
\mu \frac{du^{j}_{\nu}({\bmath{k}})}{dm} &=&
\chi^{j}_{\nu}({\bmath{k}}) v^{j}_{\nu}({\bmath{k}})
\end{eqnarray}
where we have relied on the residual symmetry of the plasma emissivity
about the magnetic field orientation, i.e.
$\eta^{j}_{\nu}({\bmath{k}}) = \eta^{j}_{\nu}({\bmath{-k}})$.
Following the usual Featrier development, we obtain the transfer
equations for the surface boundary, intermediate depth, and lower
boundary, respectively:
\begin{equation}
\label{eq:surfacebc}
\mu \chi^{-1}_{j,\nu} \frac{du^{j}_{\nu}}{dm} =
u^{j}_{\nu}-I^{j}_{\nu}({\bmath{-k}})
\end{equation}
\begin{equation}
\label{eq:second_RTE}
\mu^{2} \frac{d}{dm} \left(\chi^{-1}_{j,\nu}
\frac{du^{j}_{\nu}}{dm}\right) = \chi^{j}_{\nu} u^{j}_{\nu}-
\eta^{j}_{\nu}
\end{equation}
\begin{equation}
\label{eq:depthbc}
\frac{du^{j}_{\nu}}{dm} = \frac{3}{16}
\frac{\mathcal{F}_{cons}}{T^{3}} \chi_{R} \frac{dB_{\nu}}{dT}
\end{equation}
for the Rosseland mean opacity:
\begin{equation}
\frac{1}{\chi_{R}} = \frac{\upi}{4\sigma_{SB}T^{3}} \int
\left(\sum_{j=1}^{2} \frac{1}{\chi^{j}_{\nu}} \right)
\frac{dB_{\nu}}{dT} \d\nu
\end{equation}

At the stellar surface, the term $I^{j}_{\nu}({\bmath{-k}})$ can be
used to quantify either the intensity in mode $j$ of the incident
radiation field of an externally illuminated atmosphere, or a small
correction to the intensity field to account for the finite optical
depth of the surface stratum in a numerical calculation
$(0<\tau^{j}_{\nu}\ll1)$.  For our present discussion, we will set
$I^{j}_{\nu}({\bmath{-k}})=0$.

The large optical depth in either mode to the boundary allows for
ample redistribution of the flux by (principally) thermal absorption
and reemission; in magnetized models, the boundary flux of
(\ref{eq:depthbc}) can be arbitrarily distributed in the two modes
with no effect on the result of the radiative transfer.  In strongly
magnetized plasma, the opacity introduces an additional non-trivial
angular dependence to $\chi_{R}$ in (\ref{eq:depthbc}), neglected at
the lower boundary but otherwise accounted for in the RTEs at all
other depths.  The anisotropy ultimately has no appreciable effect on
the emergent radiation field, owing to the tremendous optical depth to
the lower boundary (typically $\tau>10^{2}$), and the plasma can be
modeled as isotropic with no effect to the emergent spectrum.

\subsection{Radiative equilibrium}
The total flux gradient is found by integration of the transfer
equation (\ref{simple_RTE}) over all frequencies and angles where the
opacity and emissivity have been expanded into their respective
thermal and scattering contributions.  Using the definition of the
(coherent) scattering emissivity $j^{sc}_{i,\nu}$, and demanding that
the flux be invariant with depth, this is
\begin{equation}
\label{eq:radeqinteg}
\int_{0}^{\infty} \d\nu \int_{+\Omega} \d\Omega \left[ \sum_{j=1}^{2}
\left( \alpha^{j}_{\nu} u^{j}_{\nu} - j^{th}_{j,\nu} \right) \right] =
0
\end{equation}
Equation (\ref{eq:radeqinteg}) is a necessary condition for radiative
equilibrium, but is not by itself sufficient to constrain the
\emph{magnitude} of the flux --- the thermal balance prescribed here
does not relate continuous variations in flux with depth.  When
simultaneously integrated with the system of radiative transfer
equations (RTEs), where the total conserved flux is specified in the
depth boundary (\ref{eq:depthbc}), equation (\ref{eq:radeqinteg})
provides a robust constraint.  To directly calculate the flux
$\mathcal{F}_{\nu}$ we evaluate the conjugate field $v_{\mu\nu}^{j}$
from either the Feautrier definitions at the surface or from eqn
(\ref{eq:Feaut_eqn_u}) at any other depth:
\begin{equation}
\label{eq:curlyF}
\mathcal{F}_{\nu} = 2 \int_{0}^{2\upi} \int_{0}^{1} \mu \sum_{j=1}^{2}
v_{\mu\nu}^{j}\,\d\mu\,\d\phi
\end{equation}

\subsection{Hydrostatic equilibrium}
Our model of the neutron star atmosphere will neglect bulk motion of
the plasma, and will therefore assume a hydrostatic structure.  The
total pressure at any depth of the atmosphere is the sum of
contributions from the ideal gas pressure, the radiation field, and
non-ideal effects owing to interactions between plasma constituents:
\begin{equation}
P = P_{ideal}+ P_{non-ideal}+ P_{rad}
\end{equation}
The ideal gas pressure, including electron degeneracy, is
\begin{equation}
P_{ideal} = NkT + n_{e}(y_{p}-1)kT
\end{equation}
where $N$ is the total number density of the plasma constituents, and
the parameter $y_{p}$ is the ratio of Fermi integrals $I_{p}(\alpha)$,
$y_{p}=I_{p}/(p I_{p-1})$.  For unmagnetized plasma, $p=3/2$, and in
quantizing magnetic fields, $p=1/2$.  The electron density is an
independent model variable in our calculation; to ensure a
self-consistent equation of state, we compare the value of $n_{e}$ of
the current iterate to that given by \citep{Clayton,Potekh99}
\begin{equation}
\label{electron_densities}
n_{e} = \left\{
\begin{array}{ll}
\frac{4}{\sqrt{\upi}\lambda_{e}^{3}} I_{1/2} & \ell>1 \\
\frac{1}{\sqrt{\upi}\lambda_{e}} \left(\frac{eB}{hc}\right) I_{-1/2} &
\ell\la 1
\end{array} \right.
\end{equation}
where $\lambda_{e}=(2\upi\beta\hbar^{2}/m_{e}kT)^{1/2}$ is the
electron thermal wavelength, and the parameter $\ell$ is
defined in (\ref{eq:landaupops}).  The electron degeneracy parameter
$\alpha$ is evaluated from $n_{e}$ either by inversion of the Fermi
integral \citep{Antia93} or by a numerical root search, from which the
degeneracy pressure is calculated.  In strongly quantizing magnetic
fields, the phase space volume occupied by the electron distribution
is much smaller than that of a weakly magnetized plasma, tempering the
onset of degeneracy which occurs at higher densities than for
unmagnetized plasma.  At intermediate magnetic field strengths, the
degeneracy contribution to the gas pressure should be evaluated by a
sum over occupied Landau states.  In practice, degeneracy pressure
contributes less than $\sim 4\%$ at even the deepest model strata, and
has no discernible influence on the emergent spectrum.  In our models
the degeneracy pressure at finite $B\la 10^{10}$ G is approximated by
the zero field limit.

Non-ideal contributions to the pressure include the Coulomb
interaction of the ionized plasma and, for an incompletely ionized
plasma, the atom-atom and atom-ion interactions in the neutral plasma
component.  The latter are neglected in our simplified EOS; for the
former, we use the Debye-H\"uckel approximation in both magnetized and
unmagnetized plasma \citep{Clayton}:
\begin{equation}
\label{eq:DebyeHuckel}
P_{Coul} = -\frac{2e^{3}}{3}\left(\frac{2\upi}{kT}\right)^{1/2} (\xi
n_{e})^{3/2}
\end{equation}
where $\xi$ is a compositional parameter $\mathcal{O}(1)$.  This
approximation breaks down when the Coulomb energy at the mean
electron-ion separation becomes comparable to the local thermal
energy, which can occur at depth in models which are both strongly
magnetized and very cool.

The radiation pressure is formally a tensor quantity but usually taken
to be isotropic in conventional atmosphere models.  Finite
anisotropies to the pressure tensor are induced in oblique magnetic
fields; however, as we shall see from the discussion of model results,
the radiation force plays only a minor role in the total pressure
structure of the neutron star atmosphere (this is especially true in
the region of photon decoupling) and we will consider only the
$\hat{z}\hat{z}$ component of the pressure tensor regardless of field
orientation in these calculations.  In CGS units this scalar pressure
is:
\begin{equation}
P_{rad} = \frac{4\upi}{c} \int_{0}^{\infty} \d\nu \int_{+\Omega}
\d\Omega \,\mu^{2} \sum_{j=1}^{2} u^{j}_{\mu\nu}
\end{equation}

The fiducial depth in our atmosphere model is the Lagrange variable
$m\ \rmn{(g\:cm^{-2})}$, in terms of which the total pressure gradient
is:
\begin{equation}
\label{eq:dP_dg}
\frac{dP}{dm} = g_{s}
\end{equation}
Eqn (\ref{eq:dP_dg}) is to be integrated simultaneously with the RTE
system and other constraints.  In the limit of $m_{0}\rightarrow0$ the
total pressure will equal that of the emergent radiation field; in
numerical calculations it is not possible to realize this limit, and
the pressure $P(m_{0})$ will retain finite contributions from each
source, dominated by the gas pressure $P_{g}$.  At the surface, we
require $P(m_{0}) = g_{s} m_{0}$.

Finally, the system of equations is closed by the EOS.  In the limit
of complete ionization, the statistical equilibrium for a
light element plasma is particularly simple: we require only that the
local charge distribution of the model be neutral, and that particle
number be conserved everywhere:
\[
n_{e} = \sum_{s\neq e} Z_{s} n_{s}, \qquad N = \sum_{s} n_{s}
\]

Although convective transport is neglected in our prescription for
energy transport and flux conservation in the stellar atmosphere,
conditions amenable to convection may arise in the resultant
atmospheric structure, and the validity of our non-convective model
hypothesis should be evaluated in the final result.  The onset of
convective instability is estimated by the Schwarzschild criterion,
generalized to include the effects of radiation pressure, and of the
neutral species gradient in unmagnetized atmospheres:
\begin{equation}
\nabla_{R} \equiv \left(\frac{d\ln{T}}{d\ln{P}}\right)_{R} >
\left(\frac{d\ln{T}}{d\ln{P}}\right)_{A} \equiv \nabla_{A}
\end{equation}
The radiative gradient $\nabla_{R}$ is evaluated from the model $T(P)$
by numeric differentiation and $\nabla_{A}$ is calculated as
described by \citet{Krishna61}.  This measure of the
adiabatic gradient does not include the effect of Ohmic dissipation in
magnetized plasma, which tends to improve the stability of the plasma
against the onset of convection.  None of the models presented in
\S\ref{Sec:Applications} show evidence of convective instability;
\citet{Zavlin95b} find that convection occurs in light element
atmospheres only for exceptionally cool models, $\Teff\la 5\times
10^{4}\ \mathrm{K}$.

\section{Solution technique}
\label{Sec:Solution}
Apart from the strategy used to refine the model, successful
convergence will often depend critically upon the properties of the
trial or seed solution, and on the distribution of points on the
discrete grids over continuous model dimensions, e.g. energy, angle
and depth.

\subsection{Seed thermal structure}
\label{Sec:Solution:Seed}
We use a power-law prescription for the plasma conductivity assuming
free-free absorption is the dominant source of opacity, i.e. the
Kramers opacity \citep{Clayton}.  This simplified model has two
functions: first, to suggest reasonable bounds for the depth variable
$m$ which will cover the range of photospheric depth for a prescribed
range of photon energies; and second, to construct an initial
temperature profile which crudely approximates radiative equilibrium.
Derivations of the mass, temperature and optical depth relations for
this model proceed from the discussion in \citet{HH98a}.

For the monochromatic optical depth $\tau$ to photons of energy $E$
(keV), the equivalent mass column density $(g\ cm^{-2})$ in this model
is given by
\begin{equation}
\label{eq:moftau}
m_{E}(\tau) = 2.74\times10^{-2} T_{\mathit{eff}}^{3/7} \left( E^{3}
\tau \right)^{17/28} \left(\mu g_{s,14}\right)^{-1/2}
\end{equation}
Here, $g_{s,14}\equiv g_{s}/10^{14}$ and $\mu$ is the mean molecular
weight of the plasma, which can be defined in some average sense for
H/He mixtures with compositional gradients.  The required depth range
is therefore approximately bounded by the range of photon energies
considered in the model.  For the NS spectra under consideration, we
take for this energy range $10^{-3}-10$ keV.  The Kramers model
underestimates the surface temperature of the converged result, and
generally overestimates the opacity of the plasma in the outermost
strata.  Consequently, it is sufficient to define the surface boundary
by evaluating (\ref{eq:moftau}) for the lowest model energy at
$\tau=0.01$, and the lower boundary by evaluating (\ref{eq:moftau})
for the highest model energy at $\tau=1.0$.  For models with
$\log{B}\geq11.0$ the lower depth boundary is multiplied by
$B_{11}\equiv B/10^{11}\rmn{G}$ to account for the reduced opacity at
X-ray energies in the extraordinary polarization mode.  The total
optical depth range for a converged model is typically $\tau\sim
10^{-5}-10^{5}$.  The temperature profile derived from the power-law
model is
\begin{equation}
\label{eq:Tofm}
T(m) \simeq 56 \left(\mu g_{s} T_{\mathit{eff}}^{4}
m^{2}\right)^{2/17}
\end{equation}

The initial density profiles can be approximated by assuming only
ideal gas pressure contributes to the total.

\subsection{Mesh generation}
\label{Sec:Solution:Grid}
The rate and quality of model convergence for a particular choice of
parameters are sensitive to the choice of discretization of the
continuous model dimensions, and will often require the careful
distribution of grid points in depth $m$, photon energies $E$ and
trajectories through the plasma $(\mu,\phi)$.  The current numerical
implementation requires the angle-energy mesh to remain fixed during
the calculation.  We expect that a reliable solution should result
from a mesh which is designed to capture significant features of the
plasma opacity with respect to these dimensions.

\subsubsection{Frequency mesh}
For most NS spectra we consider the energy range $E = 10^{-3} - 10$
keV which provides sufficient coverage of the spectrum from optical
through soft X-ray energies for a broad range of $\Teff$.  The
continuum frequency mesh is allocated by dividing the total energy
range into logarithmic subintervals, each populated with equal numbers
of mesh points.  Any of these intervals may be further subdivided to
accommodate an atomic or cyclotron feature, in which case the
continuum interval is split into smaller intervals with mesh point
densities proportional to their widths.  We typically use $6-8$ points
per $\log{E}$ to describe the continuum, with an additional $NC=4-8$
points cast in each cyclotron resonance.

\subsubsection{Angle mesh}
Several strategies for prescribing the angle mesh may be considered,
and we take advantage of symmetries about the magnetic field when
available.  The radiative transfer equations in terms of the conjugate
Feautrier variables require definition of the outgoing photon
trajectories $\mu=\left(0-1\right]$.

For weakly magnetized plasma $(\log{B}\la10)$, the anisotropy imposed
on the emergent radiation field is mild at X-ray energies.  In this
nearly isotropic plasma, we adopt a simple
Gauss-Legendre or Gauss-Radau distribution in the viewing latitude
$\mu$; the latter distribution is constructed by holding the $\mu=1$
node fixed in the Legendre polynomial.  Model results calculated in
the isotropic limit which include the normal LOS do not differ
substantially from those where this trajectory is excluded.  In this
regime, as few as 5 points can be used to describe the angular
intensity distribution.

Departures from the isotropic plasma response affect the X-ray spectra
when $\log{B}\ga 10$, where large variations in the angular dependence
of the opacity allow for photon decoupling from broad range of
plasma temperature.  For $u_{e}\ga5.3$ the geometric factor $G_{X}$
acquires a maximum at $\theta_{p}$ roughly delineating a bimodal
angular distribution of radiation: a pencil beam component directed
along the magnetic field lines, and a broader fan component at more oblique
angles.  The angular width of the pencil component declines with
increasing $u_{e}$ after achieving a maximum near $u_{e}\simeq 11$;
resonances in the polarization vectors (ionic or vacuum) also have
large opening angles, but the pencil beam transports little flux in NS
models with $B\ga10^{13}\ \rmn{G}$.  It is desirable to prescribe a
fixed angle distribution which samples the dependence across the
entire spectrum of model energies for a given magnetic field strength,
and which emphasizes the transition between the two components.  These
\emph{pencil weighted} schemes distribute mesh points on intervals of
$\log{(1-\mu)}$ where $(1-\mu)_{min}$ is generally $10^{-4}$.
Good resolution of the pencil and fan emission components can be
achieved with as few as 10-16 points.  As in the isotropic case, the
normal sighting can be included as desired.

\emph{Oblique fields \& the azimuthal mesh ---}\ For normal magnetic
field vectors, the radiation field is azimuthally symmetric models and
degenerate in the $\phi$ dimension; for this, we use a single mesh
point at $\cos{\phi}=1$.  Weakly magnetized models can also be
computed with sufficient accuracy in this approximation.  Oblique
magnetic fields $B\ga10^{10}\ \rmn{G}$ clearly break the azimuthal
symmetry and for generally oblique magnetic field inclinations we are
obliged to consider $\phi:0-\upi/2$ with emphasis on the pencil width
about the field axis $(\theta_{b},\phi=0)$.  When
$\theta_{b}=\upi/2$, the plasma response recovers partial symmetry,
and the calculation can be restricted to $\phi:0-\upi/4$, with the
remainder of the plane mapping into this range.  The direction
$\phi=0$ is defined by the projection of the (oblique) magnetic field
vector onto the plane of the atmosphere
$(\vec{x}=\hat{b}-\hat{n}\cos{\theta_{b}})$.  For generally oblique
magnetic fields, we use $5-8$ azimuth points in the following
overlapping distributions: (i) two nodes are assigned at $\cos\phi=-1,1$
and (ii) the remainder distributed logarithmically in the quadrant
$\cos\phi=[0,1]$ similarly to the pencil weighted scheme described
above.  Fully orthogonal models $(b=\upi/2)$ can be calculated with as
few as 4 azimuthal points, and for these we choose a Gauss-Lobotto 
distribution with
anchored nodes at $\cos\phi=[0,1]$. Regardless of their distribution, we
calculate the angle the trajectory and the magnetic field for each
angle pair $(\mu,\phi):$
\begin{equation}
\cos{\theta_{bk}}=\cos{\phi}\left(\sin{\theta_{b}}\sin{\theta_{k}}
+\cos{\theta_{b}}\right)
\end{equation}
for use when evaluating the geometric corrections to the plasma
opacity detailed in \S\ref{Sec:Opacity}.  Model results are invariant
under reversal of the field direction.

\subsubsection{Depth mesh}
In general, we compute the atmosphere models on a fixed depth mesh
with 10 points per decade for a total of $90-130$ points.  Models
including a strong magnetic field, with non-trivial vacuum
contributions to the opacity, often require special care; experience
has shown that these models benefit from an additional depth mesh
component which adaptively traces the coupling of density with photon
energy near the vacuum resonance (e.g. eqn
(\ref{eq:crit_vac_plas_density}), see also \citet{HoLai03}).  These
adaptive depth points are found by interpolating the current density
profile $n_{e}$ to find the critical density for vacuum resonance
formation at each model energy from (\ref{eq:crit_vac_plas_density}),
and ignoring photon energies which do not see the resonance on the
current range of $n_{e}$ described by the model.  A single adaptive
point is assigned to the ``core'' of the vacuum resonance; additional
points can be symmetrically distributed at $\rho_{res}(1 \pm n
\epsilon)$, $n=1,2\ldots$, for $\epsilon\ll1$ but this is not
required.  All independent model variables $(T,n,u)$ are then
interpolated for each adaptive depth point from the current model
solution.  In successive iterates, the previous adaptive mesh is
discarded after interpolation of the current iterate's adaptive
points.

\subsection{Formal radiative transfer calculation}
\label{Sec:Solution:Formal}
For a given atmospheric structure $T(m),n_{s}(m)$ we calculate the
opacities and thermal emissivities at each strata of the model.  The
radiative transfer equations (\ref{eq:surfacebc}-\ref{eq:depthbc}) are
transcribed to a simple differencing scheme; the discretized RTE at
intermediate depth for radiation of a particular energy, mode and
trajectory (with these indices suppressed) is:
\begin{eqnarray}
\label{eq:RTE_1}
\nonumber \lefteqn{ -\mu^{2} D_{-}u_{d-1} + \left[\mu^{2} \left(
D_{-}+D_{+}\right) + \overline{\Delta m}\chi \right] u_{d} - \mu^{2}
D_{+}u_{d+1} } \\ & & = \overline{\Delta m}(\eta^{th}+\eta^{sc})
\end{eqnarray}
and the surface and depth boundary equations are
\begin{eqnarray}
\label{eq:RTE_2}
\nonumber \lefteqn{ -\frac{\mu^{2}}{2} D_{+}
u_{2}+\left[\frac{\mu^{2}}{2} D_{+}+\mu+\frac{1}{2} \Delta m_{+}\chi
\right] u_{1} } \\ & & = \mu I^{-} +\frac{1}{2} \Delta
m_{+}(\eta^{th}+\eta^{sc}) \\
\label{eq:RTE_3}
\nonumber \lefteqn{ -\frac{\mu^{2}}{2} D_{-}
u_{ND-1}+\left[\frac{\mu^{2}}{2} D_{-}+\frac{1}{2} \Delta m_{-}\chi
\right] u_{ND} } \\ & & = 3 \mu^{2}
\frac{T_{e}^{4}}{16}\chi^{-1}\chi_{R}\frac{dB}{dT} T^{-3} +\frac{1}{2}
\Delta m_{-}(\eta^{th}+\eta^{sc})
\end{eqnarray}
for $d=2\ldots ND-1$.  The differenced boundary equations
(\ref{eq:RTE_2}-\ref{eq:RTE_3}) have been promoted to second order
accuracy by substitution of (\ref{eq:second_RTE}); higher order
accuracy can be obtained by evaluating the gradient of the emissivity
\citep{Auer76}, but does not substantially improve our results.  The
depth intervals are $\Delta m_{\pm} = \left|m_{d\pm1}-m_{d}\right|$,
$\overline{\Delta m} \equiv \Delta m_{+}+\Delta m_{-}$ and the
differencing coefficients $D$ are \citep{AHH77}:
\[
\label{eq:diffcoeffs}
D_{\pm} \equiv \frac{\left(
\chi^{-1}_{d\pm1}+\chi^{-1}_{d}\right)}{\Delta m_{\pm}}
\]

The scattering integrals are expanded as a sum over incoming photon
modes and trajectories which explicitly couples radiation in the two
polarizations states, and integrates the scattering emissivity
\textit{in-situ}:
\[
\eta_{i}^{sc}(k) = \frac{n_{e}}{\rho} \sum_{k'} w_{k'} \sum_{j=1}^{2}
u^{j}(k') \left(\frac{d\sigma_{ji}^{k'k}}{d\Omega'}\right)
\]
The angular integration weights $w_{k}$ are the product of the
respective $\mu$ and $\phi$ Gaussian weights corresponding to the
trajectories $k$.  Rendering integrations of the total opacity and
emissivities on the fixed angle mesh accurately preserves both thermal
balance and the conservative scattering processes.

For a particular model energy $E$, the coefficients of the vector
$u_{d,E}=(u_{1}^{j}\ldots u_{NA}^{j})_{d,E}$ from
(\ref{eq:RTE_1}-\ref{eq:RTE_3}) are written into matrix coefficients
\begin{equation}
\label{eq:RTEsys}
-A_{d,E} u_{d-1,E} + B_{d,E} u_{d,E} - C_{d,E} u_{d+1,E} = K_{d,E}
\end{equation}
which are generally full but diagonally dominant; the $K_{d}$ are the
thermal emissivity factors and boundary terms from eqns
(\ref{eq:RTE_2}) and (\ref{eq:RTE_3}).  For coherent scattering,
radiation streams of different energies are not coupled and, per
frequency, the RTE system is simply expressed in this block
tridiagonal form for mode-trajectory pairs organized by depth which
can be solved by standard elimination techniques.  Matrix inverses are
never explicitly evaluated -- instead the common matrix terms in the
projection operators are \textit{LU} factored and the intermediate vectors
and matrix operators then found by back-substitution.  All algebraic
reductions in our code are made using standard \textit{BLAS} and
\textit{LAPACK} routines.

The resultant radiation field, while consistent with the prescribed
atmospheric structure, cannot be expected to satisfy flux
conservation.  The trial solution will therefore be iteratively
refined by deducing corrections to each model variable such that the
entire system of RTEs and constraints are driven towards a convergent
solution.  This proceeds by complete linearization (CL).

\subsection{CL algebra}
\label{Sec:Solution:CL}
The tridiagonal organization of the formal RTE system in the previous
section provides the algebraic basis for extension to the full
iterative problem in the complete-linearization technique
\citep{AuerMihalas3}, although the detailed organization of the
coefficients is distinct from that of the formal transfer calculation.
The constrained RTE system is again transcribed to a numeric grid in a
simple differencing scheme, but now \textit{linear perturbations} to
the radiation components and physical fields are the independent model
variables.  The correction to any model variable will be associated
with a linearized equation: either an RTE (for $u$) or constraint (for
$T,N,n_{s}$).  The solution is represented by the vector
$\vec{\Psi}=(\Psi_{1}\ldots\Psi_{D})$, where
$\Psi_{d}=(N,T,n_{s},u^{j}_{\mu\nu})_{d}$ are formed from the current
model state.  Corrections to the model solution are found by solving
an equation of the form $O\delta\vec{\Psi}=E$ where $E$ is the error
in the current solution, and $O$ is an operator derived from
coefficients of the perturbed equations.  For brevity we sketch the
main elements of the technique; see \citet{Mihalas78} for a complete
example of complete linearization applied to a simple atmosphere.

In each model equation, the independent variables are replaced by
first order corrections, $T \rightarrow T+\delta T$, etc.  Functions
are linearized by taking partial derivatives with respect to the
independent model variables:
\begin{equation}
\label{eq:chi_deriv_expand}
\chi \rightarrow \chi + \frac{\partial\chi}{\partial T}\delta T +
\sum_{s}\frac{\partial\chi}{\partial n_{s}}\delta n_{s}
\end{equation}
and similarly for the emissivities $\eta^{th}$ and $\eta^{sc}$.  The
efficiency of the CL cycle is diminished by properties held constant
during the iteration, therefore differentiation like that in
(\ref{eq:chi_deriv_expand}) includes partials taken with
respect to the squared polarization components and the Gaunt factors.
All derivatives in the model are evaluated analytically at the current
values of the model variables; for numerical work these are most
conveniently expressed in logarithmic form.  Upon substitution of all
model variables, the linearized equations are reorganized by
collecting terms from the unperturbed equations to form the error $E$,
while coefficients of each model variable correction are grouped by
depth.  The CL coefficients of the constrained system of equations are
written into matrix coefficients
\begin{equation}
-A_{d}\delta\Psi_{d-1}+B_{d}\delta\Psi_{d}-C_{d}\delta\Psi_{d+1}=E_{d}
\end{equation}
where the matrix coefficients $A,B,C$ now contain the full spectrum of
differencing coefficients, integration weights and constants of the
linearized RTE system and its constraints per depth, and $E_{d}$ is
the error vector whose components are given by the difference between
the RTEs (constraints) for $u$ $(N,T,n_{s})$ and the exact solution
calculated from the current model state.  Coefficients of the $j$th
variable at depth $d-1$ in the $i$th equation at depth $d$ are written
into the $A_{d,ij}$ etc.  The entire system forms a block tridiagonal
matrix which is solved by \textit{LU} factorization and substitution.
The new model state is found by updating each model variable with its
linear corrections, and the system is iteratively corrected in this
fashion until the root vector is rendered within some prescribed
tolerance to be described below.  This procedure is multivariate
Newton-Raphson iteration.  Note that information (model corrections)
from a given stratum propagate over the mean free path of the
radiation field via the transfer equations, and model convergence is
global.

CL often over-steps the convergent solution in early stages of the
computation, and we limit the magnitude of the temperature corrections
to $\left|\Delta T/T\right|\leq(0.4-0.7)$ to prevent model variables
from acquiring non-physical values from ``rogue'' iterates.  When this
condition is engaged at any depth, corrections to all other fields are
discarded, the density made consistent with $g_{s}$ and the
correction-limited $T(m)$ as in \S\ref{Sec:Solution:Seed}, followed by
a formal recalculation of the radiation field
(\S\ref{Sec:Solution:Formal}).  The typical CL cycle is summarized by
(1) allocation of mesh points in all model dimensions, either from
command-line arguments to the program, or through defaults; (2) this
is followed by tabulation of the Gaunt factors and their temperature
derivatives for each model photon energy in magnetized models, over a
range of temperature $\lgTeff\pm1.5$; (3)
calculation of the trial thermal structure $T(m), n_{s}(m)$, and the
run of opacity $\chi^{j}_{\nu}$ and emissivity $\eta^{j}_{\nu}$ with
depth for all photon trajectories $\theta_{bk}$ in each mode; (4) a
formal calculation of the radiation field from these opacities
(\S\ref{Sec:Solution:Formal}); followed by (5) the CL iterates.  We
adopt as acceptable convergence criteria that $\Delta T/T<10^{-3}$ and
$\Delta F/F_{cons}<10^{-3}$ at all depths. These limits are achieved
in 5-13 iterates for typical choices of computational grid densities
and dimensions, while the convergence rate can become superlinear in
the final iterates.

\section{Model results}
\label{Sec:Applications}
We have implemented the methodology of \S\ref{Sec:Solution} in an
efficient stellar atmosphere code, and computed model atmospheres and
spectra for a range of parameters consistent with measured and
inferred properties of cooling, isolated NS.  We consider stars with
surface magnetic fields up to $10^{14}$ G, and assume an ionized
Hydrogen plasma photosphere described by the opacities formulated in
\S\ref{Sec:Opacity}.  For weak and itermediate field strengths, we
consider the range $5.5\leq\lgTeff\leq6.5$, while for the magnetar
case we consider only $6.2\leq\lgTeff\leq 6.7$.  We restrict our
discussion to models with surface gravity $\log{g_{s}}=14.38$,
corresponding to NS of canonical mass $M=1.4\ M_{\sun}$ and radius
$R=10\ \rmn{km}$.  Spectra are illustrated in the stellar surface
frame and must be redshifted by $z=0.306$ to describe the flux
measured by a distant observer. All results are derived from models
converged to the tolerance described in \S\ref{Sec:Solution:CL}.  No
smoothing is applied to figures derived from converged model results.

The principal energy scaling of the geometric corrections $G_{i}(k)$
and $G_{i}^{\mathit{ff}}(k)$ enter through the magnetic field strength
in the ratio $u_{e}=(\omega_{c,e}/\omega)^{2}$, and it is natural to
organize our discussion of model results by ranges of $B$.  This
grouping emphasizes the discriminatory properties of the stellar
spectra, but also naturally collects results appropriate for
comparison to different classes of observed sources (quiescent
X-ray binaries, radio pulsars, anomalous X-ray pulsars).  The
selection of models presented in this discussion, while not
exhaustive, is representative of typical values for observed
manifestations of NS thermal X-ray emission.

X-ray polarimetry is a potentially predictive measure of the surface
properties of cooling NS, but is of immediate interest only for
relating the model input physics to the theoretical spectra.  In our
present study, it is adequate to define the net polarization $\Pi(E)$
in terms of the partial flux in each polarization mode
(\ref{eq:curlyF}):
\[
\Pi(E) = \frac{\mathcal{F}^{X}(E)-\mathcal{F}^{O}(E)}
{\mathcal{F}^{X}(E)+\mathcal{F}^{O}(E)}
\]
The flux input to the model (\ref{eq:depthbc}) can have arbitrary
polarization content; the radiation field is efficiently redistributed
by absorption and scattering processes and the final spectrum is
independent of the input polarization fraction provided
$\tau_{E}^{base}\gg 1$.  The polarization content of the
\textit{intensities} at large depth is essentially zero, and that of
the final emergent spectrum is determined by the relative temperatures
at which the two modes decouple from the atmosphere.  Upon decoupling
from the plasma, the intensity and its angular distribution (i.e. flux
and pressure) in either mode asymptotically acquire values which are
unchanging in the higher strata.  As the gas pressure declines toward
the atmospheric boundary, the radiation pressure assumes the dominant
role; exterior to the star, the pressure is entirely that of the
radiation field.  All model calculations described here account for
the radiation pressure, but throughout the range of photon decoupling
its contribution to the total $P$ is negligible.

\subsection{Unmagnetized models $\bmath{(B\la 10^{10}\ \rmn{G})}$}
\label{Sec:Apps:Unmag}
\begin{figure}
\includegraphics[width=84mm]{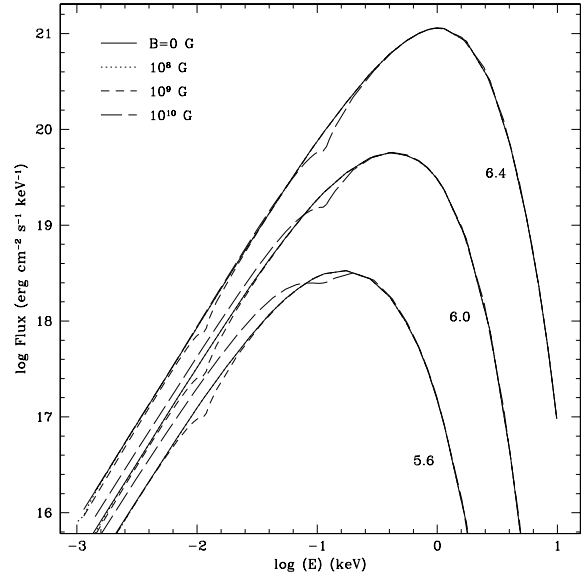}
\caption{ Total emergent flux profiles for unmagnetized Hydrogen
plasma atmosphere (solid curves), and weakly magnetized plasma
$\log{B}=8.0(1.0)10.0$ G, at three effective temperatures
$\lgTeff=5.6(0.4)6.4$.  For $\log{B}=10.0$ the cyclotron absorption
encroaches on the soft X-ray spectrum, and at lower $\Teff$ becomes
comparable to the peak emission, yielding enhanced optical and near UV
amplitudes.  With the exception of the coolest $\log{B}=10.0$
atmospheres, the model spectra satisfy a single displacement rule
$E_{\gamma}^{max}\simeq 4.7 k\Teff$.  }
\label{fig:unmag_tot_fluxes}
\end{figure}

The onset of magnetic corrections to the plasma opacity occurs where
$u_{e}\simeq 1$, and the electron cyclotron feature clearly delineates
the polarizing properties of the magnetized plasma.  Consequently, the
X-ray spectra of magnetized NS atmospheres having $B\la 10^{10}\
\rmn{G}$ are essentially indistinguishable from the zero-field case
for the same plasma composition and total flux, and may be classified
as \textit{unmagnetized} models.  The transition from classical to
quantizing magnetic field occurs within the range of $\Teff$ described
for $B=10^{9}-10^{10}\ \mathrm{G}$.  As described in
\S\ref{Sec:Opacity} our calculation of the bremsstrahlung Gaunt factor
is somewhat inaccurate for $\ell \sim 1$, and here we will assume the
zero-field $g$ for each effective temperature at these field
strengths.  Consequently, the polarization properties derived for
models in this regime are influenced only by the normal mode vector
decomposition, but permit comparison between models at different
$\Teff$ having the same field strength and guarantee consistent
calculation of $g$ at all strata of a given model.  In Figure
\ref{fig:unmag_tot_fluxes}, we compare spectra calculated in the weak
field regime for a range of \Teff and find no magnetic field
dependence for the emission $E \ga0.2\ \rm{keV}$.  For the range of
$\Teff$ considered here, the spectra in the unmagnetized field regime
satisfy a linear relation between the peak spectral energy the
effective temperature similar to the Wien displacement of the Planck
function; $E_{peak} \simeq 4.7 k\Teff$ in these unmagnetized models.

Departures from the zero field results at optical and UV wavelengths
for finite $B$ arise from the redistribution of resonantly absorbed
radiation to longer wavelengths in accordance with flux constancy in
radiative equilibrium.  The magnitude of this redistribution is
greater where the cyclotron line is approximately coincident with the
peak thermal emission (e.g. $\lgTeff=5.6$ at $10^{10}\ \rmn{G}$) owing
to both the scaling of the effective cyclotron line width and because
there is more radiation in the spectral peak which can be reprocessed.
Additional details of the influence of cyclotron energy redistribution
on visual magnitudes are presented elsewhere \citep{Lloyd03}.  More
generally, photons absorbed in a neutral plasma component is
likewise reemitted at longer wavelengths in radiative equilibrium
producing enhanced emission below the absorption feature.

\begin{figure}
\includegraphics[width=84mm]{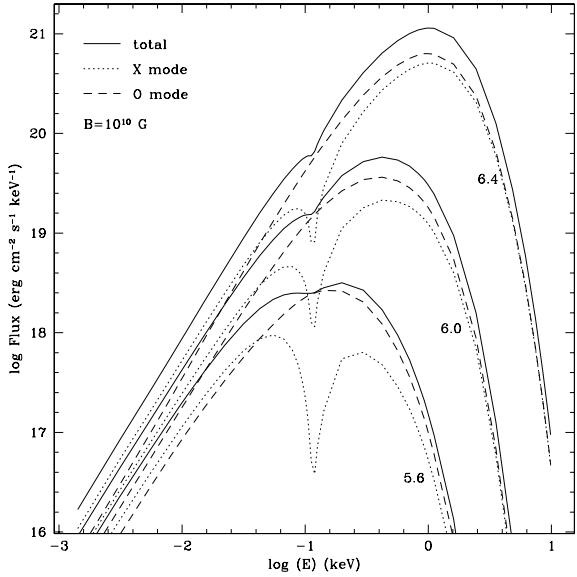}
\caption{ Spectral envelopes for $\log{B}=10.0$ at three flux
temperatures, $\lgTeff=5.6(0.4)6.4$; resonant absorption significantly
affects the spectrum where the electron cyclotron feature is nearly
coincident with peak emission energy. Only X-mode radiation is
sensitive to electron cyclotron absorption, and energy thus absorbed
is reemitted at longer wavelenths.  }
\label{fig:unmag_10_polz_fluxes}
\end{figure}

\begin{figure}
\includegraphics[width=84mm]{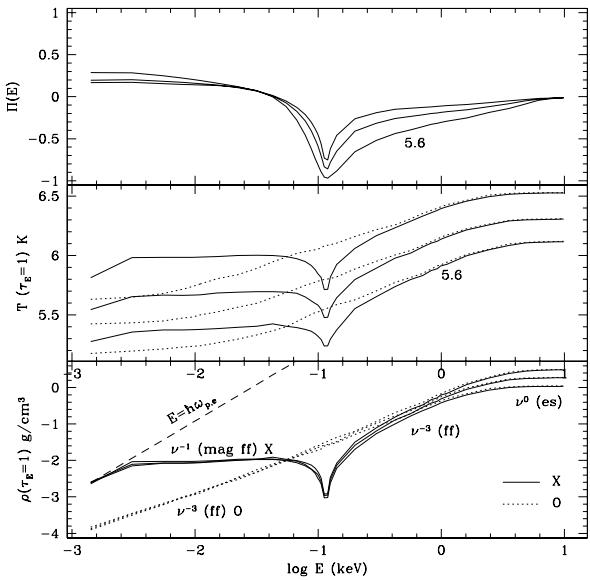}
\caption{ Polarization fraction (upper), photospheric temperature
(middle) and decoupling density (lower panel) for the model spectra of
Fig \ref{fig:unmag_10_polz_fluxes}.  Resonant absorption plays a more
significant role at cooler temperatures, where the peak thermal energy
approximately coincides with the cyclotron line.  Thomson scattering
of high energy photons is energy independent; lower energy photons
which are predominantly absorbed decouple from similar densities
regardless of \Teff. }
\label{fig:unmag_10_polz_details}
\end{figure}

Only X mode radiation is resonantly absorbed in the electron cyclotron
line, and the excess ordinary mode contribution yields a large net
polarization in the resonance.  The effective width of the cyclotron
feature is approximately scale invariant.  At optical and near UV
wavelengths the X mode photosphere is displaced to higher
temperatures, owing to the reduced opacity in this mode.  The spectrum
is largely unpolarized at energies $\omega\gg\omega_{c,e}$, for which
the opacity reverts to the anticipated unmagnetized result, although,
as the magnetic field strength approaches $10^{10}\ \rmn{G}$
$(\hbar\omega_{c,e}\simeq 0.12\ \rmn{keV})$, the unpolarized limit is
recovered only gradually at high energies.  For energies $E\ga 2.5\
\rmn{keV}$, the photosphere temperature is saturated at $T\sim
2\times10^{6}$ K where (frequency independent) electron scattering is
the principal opacity source.  The polarized spectra and plasma
decoupling properties of the $10^{10}\ \rmn{G}$ models are illustrated
for several \Teff in Figs. \ref{fig:unmag_10_polz_fluxes} and
\ref{fig:unmag_10_polz_details}.  The decoupling density
$\rho(\tau_{E})=1$ in each mode is approximately independent of flux
temperature for a given value of $B$.  Collective phenomena become
significant in the plasma response at long wavelengths, are not fully
described by the opacity formulated of \S\ref{Sec:Opacity} -- this
limitation can be seen at optical frequencies in the $10^{10}\ \rmn{G}$
models where photons decouple at $v_{e}\simeq 1$
(Fig. \ref{fig:unmag_10_polz_details}).

\begin{figure}
\includegraphics[width=84mm]{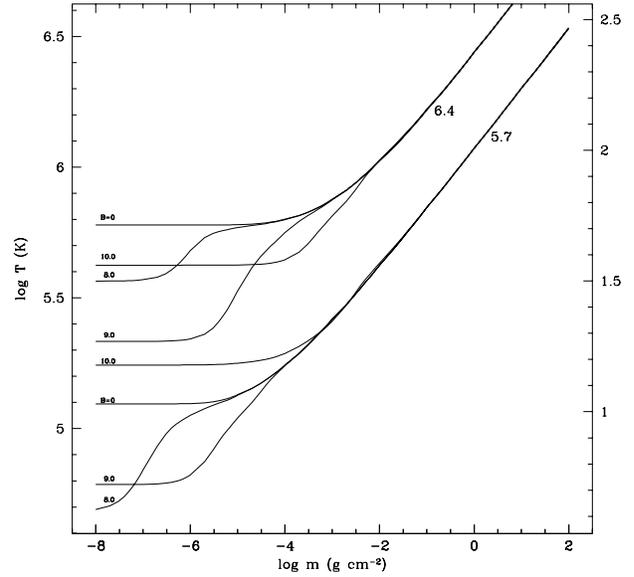}
\caption{ Temperature profiles for several converged, weakly magnetized
models, illustrating the variation in asymptotic surface temperature which
results from energy redistribution from cyclotron absorption in a
radiative equilibrium atmosphere.  Curves are labelled by $B=0$
(unmagnetized) or $\log{B}$.  At finite $B$, departures from the
unmagnetized thermal structure occur at deeper strata for increasing
field strength.  }
\label{fig:unmag_Tofm}
\end{figure}

Radiative equilibrium mitigates the influence of cyclotron absorption
by adjusting the asymptotic surface structure $T(m)$.  Figure
\ref{fig:unmag_Tofm} demonstrates the magnitude of this effect for two
values of $\Teff$.  The ordering of surface temperature $T(m_{0})$
with magnetic field differs for each value of \Teff, and the departure
from the unmagnetized temperature profile at finite $B$ occurs at
greater depth with increasing magnetic field.  At depth where only
photons with $\omega>\omega_{e,c}$ decouple, the models have identical
thermal structure.  Radiation emerging from this isotropic plasma
has an angular distribution characteristic of limb-darkening, which
is preserved for oblique magnetic fields of modest strenght.

\subsection{Intermediate fields $\bmath{(10^{11}\la B\la 10^{13}\ \rmn{G})}$}
\label{Sec:Apps:Radio}
\begin{figure}
\includegraphics[width=84mm]{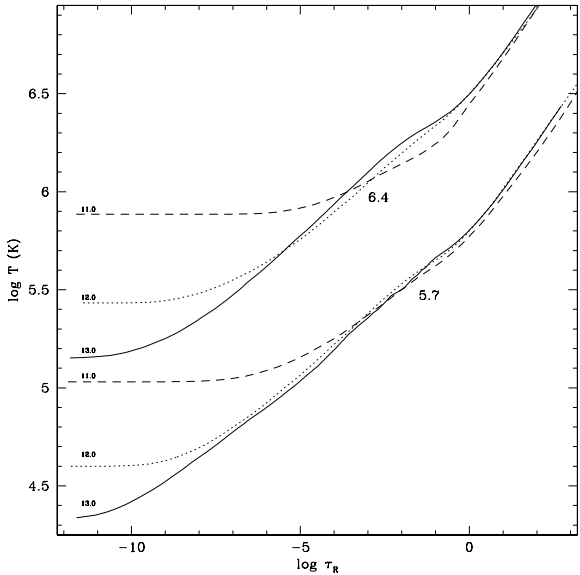}
\caption{ Equilibrium temperature profiles on the mean Rosseland depth
scale for $\lgTeff=5.7,6.4$ with intermediate magnetic fields
$\log{B}=11.0(1.0)13.0$.  As $B$ increases, a greater fraction of the
total flux is emitted in the (transparant) X mode from deep, hot
strata.  The asymptotic surface temperature declines with increasing
field strength owing to the depleted intensity in the more opaque O mode.
}
\label{fig:radio_ToftauR}
\end{figure}

In the unmagnetized plasma regime, geometric corrections to the
opacity are both weak in amplitude, and nearly independent of
propagation angle; consequently, high energy emission in weakly
magnetized models is well described by the equivalent zero field
results.  Departures from isotropic plasma response are prominent in
the X-ray spectra of models with $10^{11}\la B\la 10^{13}\ \rmn{G}$
comparable e.g. to the radio pulsars (RPs) where they are manifest in
the finite polarization $\Pi(E)$ of the spectrum and in the angular
distribution of intensities.  Polarization of the thermal radiation is
essentially complete for $B\ga 10^{12}$ G, and the spectral envelopes
of these models differ substantially from their weakly magnetized
counterparts, owing to the harder energy dependence of the X mode
opacity ($\propto \nu^{-1}$) which suppresses the amplitude of
emission in the high energy tail and shifts the peak emission energy
to lower frequencies.  Inflections in the equilibrium thermal
structure (Fig. \ref{fig:radio_ToftauR}) are related to displacement
of X mode decoupling to deeper, hotter strata than for the ordinary
mode.  In this regime, the asymptotic surface temperature declines for
increasing $B$ owing to the reduced energy density in the O mode
field.  The peak energy displacement with effective temperature is
$E_{peak} \simeq 4.0 k\Teff$, and the spectrum is dominated by
radiation component is the X mode field.  Vacuum corrections to the
plasma dielectric are negligible in these models.

\begin{figure}
\includegraphics[width=84mm]{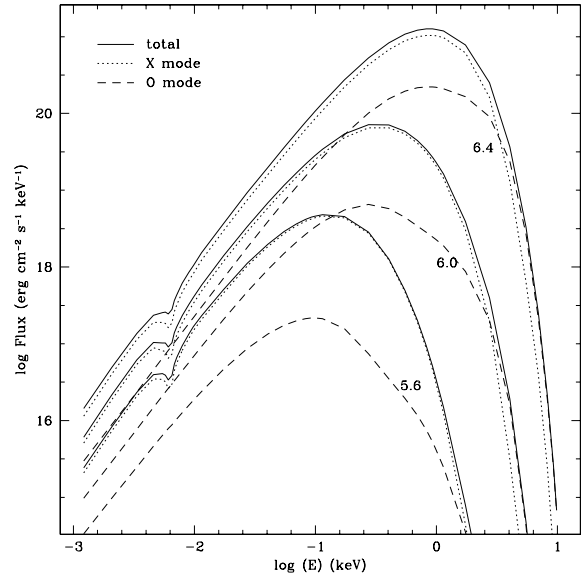}
\caption{ Model spectra for $\log{B}=12.0$ and $\lgTeff=5.6(0.4)6.4$
including the partial polarized flux contributions.  The broad electron
cyclotron line is centered off diagram at $E\sim 11.6\ \rmn{keV}$, but
its influence is seen in the relative amplitude of emission in the two
modes.  }
\label{fig:radio_12_polz_fluxes}
\end{figure}

The proton cyclotron resonance enters the optical regime at
$\log{B}\simeq 11.5$, and soft X-rays at $B\simeq 10^{13}$ G.  The
apparant depth of the feature is greater than that of the electron
feature seen in weakly magnetized models owing to the overall
depletion of O mode intensity at $\omega<\omega_{c,e}$.  The proton
line width scales with the cyclotron frequency similar to that of the
electron resonance (Fig. \ref{fig:radio_12_polz_details}).  Collective
phenomena play an increasingly important role at long wavelengths in
these models, and the energy below which the plasma attenuation
becomes significant is $E_{plas} \simeq 0.5\log{B}-7.6$.  At these
long wavelengths the specific photosphere is coincident with the
boundary $v_{e}$ owing to the large optical depth accumulated at this
boundary.  The broad electron resonance $E\sim 11.6\ \rmn{keV}$ for
models of $10^{12}\ \rmn{G}$ is not resolved in these calculations,
though its influence is obvious in the polarized spectra in Figures
\ref{fig:radio_12_polz_fluxes}- \ref{fig:radio_12_polz_details}.

\begin{figure}
\includegraphics[width=84mm]{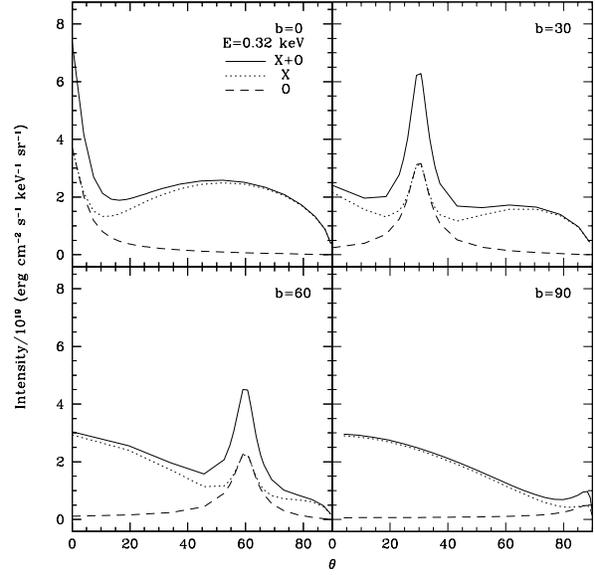}
\caption{ Angular intensity profiles near the spectral peak for
$\lgTeff=6.0$, $\log{B}=12.0$, and four choices of magnetic field
orientation.  For each curve the vectors $\vec{B},\vec{k},\vec{n}$ are
coplanar ($\phi=0$).  Intensity along the field vector declines with
increasing inclination by decoupling from lower temperatures in higher
strata.  The pencil beam has approximately equal contributions from
both modes while the fan emission is highly polarized.  }
\label{fig:radio_12_inclined_field_plane}
\end{figure}

\begin{figure}
\includegraphics[width=84mm]{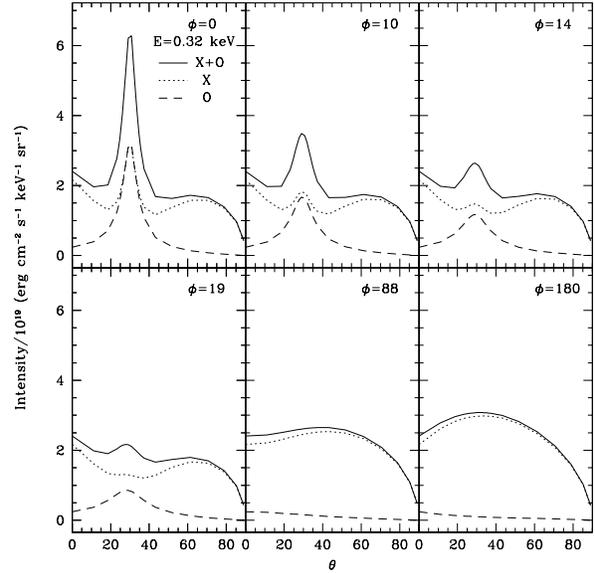}
\caption{ Angular intensity profiles for $B=10^{12}$, $\Teff=10^{6}$ and
$b=30\degr$ at a photon energy near the spectral peak for six choices
of model azimuths.  The pencil amplitude declines and acquires finite
polarization for $\phi>0$ and vanishes for $\phi\ga\theta_{p}$ and the
fan component dominates for large $\phi$.}
\label{fig:radio_12_inclined_30_per_phi}
\end{figure}

\begin{figure}
\includegraphics[width=84mm]{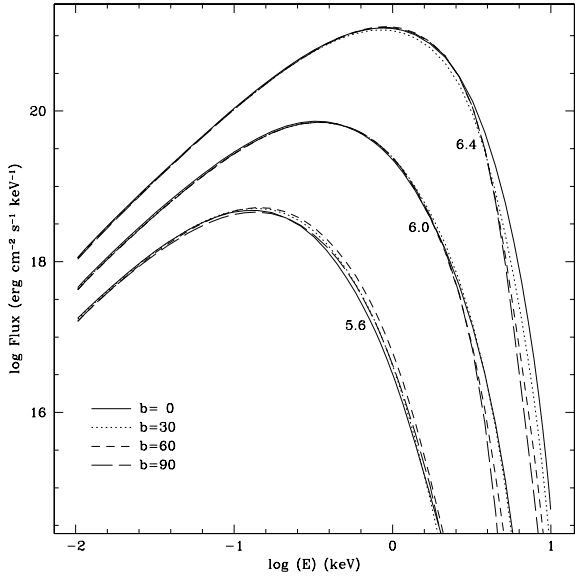}
\caption{ Total flux profiles for $\log{B}=12.0$,
$\lgTeff=5.6(0.4)6.4$ and inclinations $b=0,30,60,90$.  Spectra for
$\Teff\ga6.0$ are softer at high energies for larger field
inclinations.  The converged $T(m)$ for these models do not reveal
large variation with $b$ for a given \Teff.}
\label{fig:radio_12_inclined_fluxes}
\end{figure}

The angular intensity distributions in the intermediate field regime
acquire a pronounced pencil beam component oriented with the magnetic
field vector.  The pencil is essentially unpolarized due to the
degenerate opacity for $\theta_{b}\simeq0$ but acquires large net
polarization away from this trajectory.  The amplitude of the pencil
component declines with increasing magnetic field inclination due to
previously mentioned limb effects, but the pencil width is
approximately independent of inclination
(Fig. \ref{fig:radio_12_inclined_field_plane}).  The O mode opacity is
not magnetically depressed outside the pencil and the intensity
profile acquires a highly polarized fan-like profile dominated by X
mode emission (Fig. \ref{fig:radio_12_inclined_30_per_phi}).  Despite
the large intensity in the pencil, less than $10 \%$ of the total flux
is transported in this component at $10^{12}\ \rmn{G}$; and by
$10^{13}\ \rmn{G}$, less than $1\%$ is emitted in the pencil.  In
general, the O mode flux contribution declines with increasing field
inclination $b$, as does the total emission in the high energy tail
(Fig. \ref{fig:radio_12_inclined_fluxes}).

\subsection{Intense field limit $\bmath{(B\simeq 10^{14}\ \rmn{G})}$}
\label{Sec:Apps:Magnetar}
The abrupt X mode opacity variation induced by vacuum polarization
plays a crucial role in the equilibrium thermal structure and spectral
properties of ultramagnetized plasma atmospheres $(B\ga10^{14}$ G).
The magnetic suppression of the X mode opacities is greater in this
regime than for the RPs, and O mode intensities contribute $\la 1\%$
of the total flux.  Strictly, the precise polarization content of the
radiation field is indeterminate in the normal-mode analysis as
radiation propagates from the plasma dominated regime, through the
resonance into the vacuum dominated regime.  The shape of the spectral
envelope is substantially affected by both direct absorption within
the opaque layer, and its indirect influence on the equilibrium
thermal structure.  The displacement rules described for $B\la10^{13}\
\mathrm{G}$ is ambiguous for $10^{14}$ G owing to spectral reshaping
through vacuum effects.  The pencil component is of negligible size in
magnetar X-ray spectra and contributes $\ll1\%$ of the total flux.

\begin{figure}
\includegraphics[width=84mm]{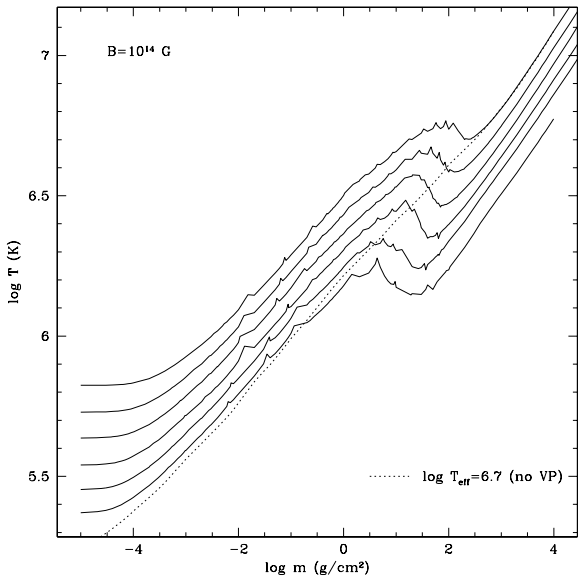}
\caption{ Thermal structure of the magnetar models $B=10^{14}$ G for a
range of effective temperatures $\lgTeff=6.2(0.1)6.7$.  The effect of
the vacuum resonance on the transparant X mode is clearly seen in the
form of the heating layer.  Small ``jumps'' in the temperature
profiles arise from abrupt variation in opacity at the adaptive depth
points, particularly for energies $E\ga 1 $ keV.  The dashed profile
is that of the converged model for pure plasma at $\lgTeff=6.7$ and
$B=10^{14}\ \rmn{G}$. }
\label{fig:magnetar_14_Tofm}
\end{figure}

Energy absorbed from the X mode field in the vacuum resonance heats
the local plasma.  There is substantial variation in the precise depth
at which the absorption will occur across the soft X-ray band
(\ref{eq:crit_vac_plas_density}) and the net effect is the formation
of an opaque ``heating layer'' in the atmosphere (Figure
\ref{fig:magnetar_14_Tofm}).  The influence of this absorption on the
emergent spectrum depends on the proximity of the absorbing layer to
the X mode decoupling layer.  Both the density of resonant absorption
and the amplitude of the X mode opacity resonance induced by vacuum
corrections \textit{decline} with photon energy. Radiation of
sufficiently low frequency will have only modest interaction with the
opaque layer and will retain the plasma dominated photosphere
$(\tau_{E}^{vp}\ll 1)$.  This may be contrasted with radiation which
decouples from within the opaque layer because of the large optical
depth accumulated at $\rho_{res}$.  For the $10^{14}$ G models
illustrated, this transition occurs at $\sim 1.5-3.0$ keV.  Figure
(\ref{fig:magnetar_14_unitrho}) shows the decoupling density of these
strongly magnetized models.  At energies $\ga 3$ keV, the X mode
radiation decouples from somewhat cooler strata than would be found
were VP effects neglected; no radiation decouples from the relatively
tenuous vacuum dominated plasma $(\tau_{E}^{vp}\gg 1)$ and the
response properties characteristic of this regime have no influence on
either the polarization or angular distribution of the spectra (see
\S\ref{Sec:VP}).

\begin{figure}
\includegraphics[width=84mm]{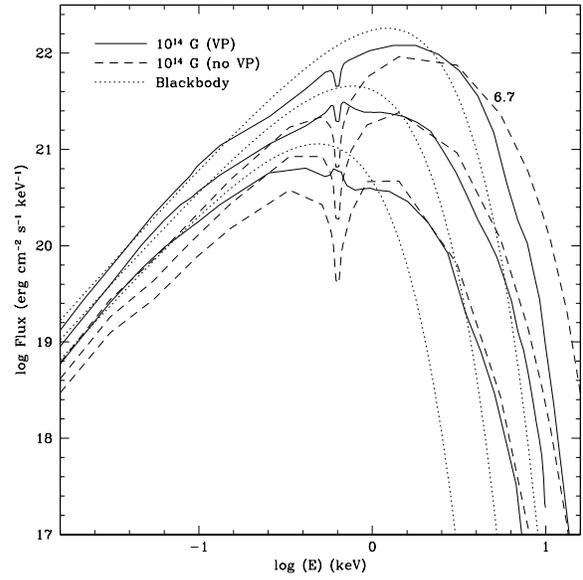}
\caption{ Total flux profiles for $\lgTeff=6.3(0.2)6.7$ at $10^{14}$
G, comparing the vacuum polarized models (solid curves) to their
ordinary plasma equivalents (long dash).  Reemission of thermal
radiation absorbed in the opaque layer (vacuum resonance) effectively
``fills in'' the proton cyclotron line in hot models, and inverts the
line profile in relatively cool models.  }
\label{fig:magnetar_14_BB_fluxes}
\end{figure}

\begin{figure}
\includegraphics[width=84mm]{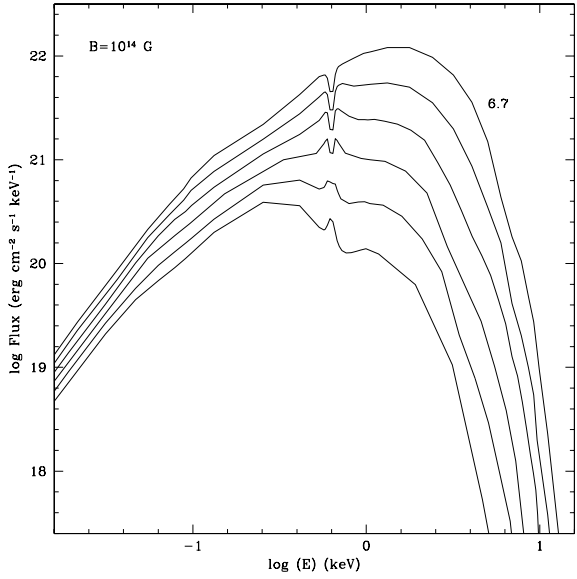}
\caption{ The flux profiles for the magnetar models $B=10^{14}$ G for
the effective temperatures $\lgTeff=6.2(0.1)6.7$.  The width of the
proton cyclotron line is reduced by the incidental effects of vacuum
polarization on the atmospheric structure (Fig
\ref{fig:magnetar_14_Tofm}).  This phenomenon is ultimately
responsible for line ``inversion'' in cooler models.  }
\label{fig:magnetar_14_fluxes}
\end{figure}

Thermal absorption in the opaque heating layer has an ancillary effect
on the spectrum at long wavelengths which decouple below the layer,
where the flux amplitude is enhanced by the re-radiated emission.  The
formation depth of the proton cyclotron line becomes comparable to the
depth of the heating layer, which reduces the effective width of the
proton cyclotron line in the $10^{14}$ G models.  For models
$\lgTeff\ga6.5$ the cyclotron absorption line is preserved but
narrower than expected from the equivalent model without vacuum
corrections (Fig. \ref{fig:magnetar_14_BB_fluxes}).  For cooler \Teff,
the formation depth of the line coincides with larger $T$ than for
neighboring frequencies, leading to an apparent emission feature.
Regardless of \Teff, the reduced effective width of the line results
from enhanced thermal emission from the opaque layer.  That this
tranformation occurs over a range of just 2.5 \Teff\ suggests that
this feature may be of diagnostic value in thermal spectra of some
anomalous X-ray pulsars.  Figure (\ref{fig:magnetar_14_fluxes})
summarizes the evolution in cyclotron temperature with $\Teff$. 

\section{Discussion}
\label{Sec:Discussion}
Observations of thermal spectra from cooling isolated NSs provide a
direct probe of their surface properties, assuming these data can be
adequately interpreted with a realistic spectral model.  The surface
magnetic field strength can be measured through the detection of
cyclotron features.  Strong atomic transitions constrain the ratio
$M/R$, while an additional measure of the gravitational potential
(e.g. through line profiles) could restrict $M$ and $R$ individually
\citep{Paerels97}; line broadening mechanisms and variation of surface
properties pose additional difficulty for interpretation of these
properties.  With the exception of 1E 1207.4-5209 for which the nature
of absorption features is presently uncertain, the failure to measure
\textit{any} spectral features in the \textit{Chandra} and
\textit{XMM} observations of thermally emitting NS is notorious.
Non-detection of features in NS surface emission tends to favor light
element light element atmosphere models; the Fe atmospheres of
\citet{RRM97} for example predict strong absorption edges at soft
X-ray energies which should be unambiguous in the data.  The absence
of cyclotron features may exclude ranges of surface field strength
but, for well magnetized stars, such features can be smoothed by
realistic distributions of the surface temperature and field
\citep{Zane01}.  Measurement of the integrated net polarization at
X-ray energies for cooling, isolated NS may provide enough information
to deduce the surface magnetic field strength where spectral features
remain unresolved or are destroyed by other processes
\citep{Pavlov00,Heyl03a}.

We have presented a flexible, efficient and robust method for
computation of light element thermal NS spectra and atmospheric
structures.  The radiative processes in these models are the simplest
required to describe the X-ray properties of cooling light element NS
atmospheres for a broad range of stellar parameters.  A number of
refinements to the model physics and distinguishing elements of the
present work are described below.

\textit{Ionization equilibrium ---} In strong magnetic fields the
ground state binding energy of stationary Hydrogen atoms grows
asymptotically with $\sim \ln^{2} B_{c}$ Ryd where $B_{c} = B/2.35
\times 10^{9}$ G \citep{Louden59}, and bound state transitions may
become important to the X-ray spectrum of NS at $10^{12}$ G.  Motional
effects of the atoms within strong magnetic fields leads to the
formation of decentered states, and yields substantially altered cross
sections \citep{Potekh98}.  \citet{Pavlov95} have shown that even a
modest neutral plasma component can alter the spectrum owing to
bound-free transitions from these tightly bound states.  Moreover,
pressure destruction is less effective in strong magnetic fields owing
to the reduce phase space volume accessible to the electron
distribution.  An occupation probability (OP) formalism has been
studied in detail for both magnetized \citet{Potekh99} and
unmagnetized Hydrogen plasma \citep{Potekh96}.  This free-energy
picture is parametrized by the binding energies and mean atomic sizes,
and provides a natural equation of state governing the pressure
destruction of weakly bound states and truncation of the partition
function.  Steady progress has been make in the EOS for magnetized
hydrogen, and a recent report describs an effort to integrate the
ionization equilibrium of \citet{Potekh02} with the Lucy-Uns\"old
temperature correction scheme \citep{Ho02b}.

It is more challenging to incorporate non-LTE processes to the
statistical equilibrium, by which the bound state distributions are
determined through detailed balance of the collisional and radiative
rate coefficients.  The radiative rates can be evaluated by
straightforward integration given (e.g.)  the bound-free cross
sections and the state of the radiation field.  The collisional
transition rates must also be specified, unlike LTE in which it is
sufficient to consider ratios of these rates and which are given by
Boltzmann factors; these rates, and preferably their derivatives, must
be computed or otherwise tabulated for the spectrum of bound states
and continuum.  The CL technique is able to respond to neutral or
incompletely ionized component better than Lucy-Uns\"old upon
simultaneous integration of the statistical equilibrium and radiative
transfer, but may require adaptive depth points around the transition
zones between successive ionization states or near line formation
depths.  Much simpler would be inclusion of OPAL opacities in the
present code (see e.g. \citet{Werner00} or \citet{Gaensicke02}) but
these tabulations are not appropriate for strongly magnetized
atmospheres.  The ionized model calculations described in this paper
underutilize the advantages afforded by the CL method.

\begin{figure}
\includegraphics[width=84mm]{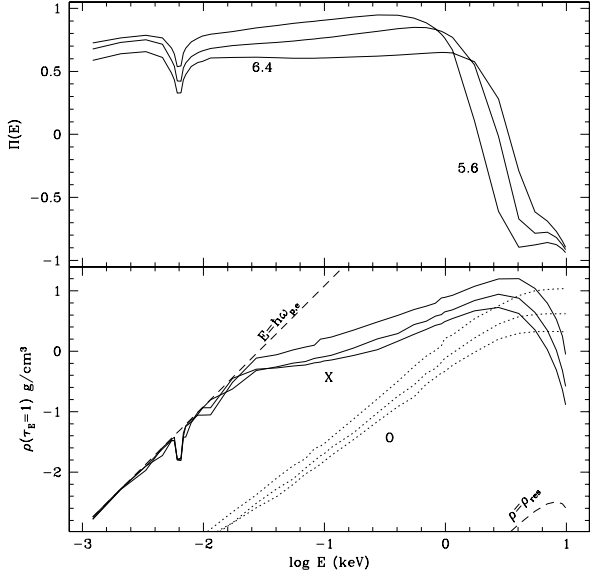}
\caption{ Polarization and decoupling density for the model spectra of
Fig (\ref{fig:radio_12_polz_fluxes}).  Radiation near the electron
feature have large negative $\Pi(E)$ owing to the relatively large O
mode amplitude (upper panel).  The proton line demonstrates modest
variation in net polarization.  The $\nu^{-1}$ dependence of the X
mode opacity is more apparant than for the $10^{10}\ \rmn{G}$ models
(lower panel).  Models with higher \Teff decouple at greater $\rho$.
Measurement of $\Pi(E)$ in two bands may be sufficient to diagnose $\Teff$ 
for surface fields $B\simeq10^{11}-10^{13}\ \rmn{G}$.
}
\label{fig:radio_12_polz_details}
\end{figure}

\begin{figure}
\includegraphics[width=84mm]{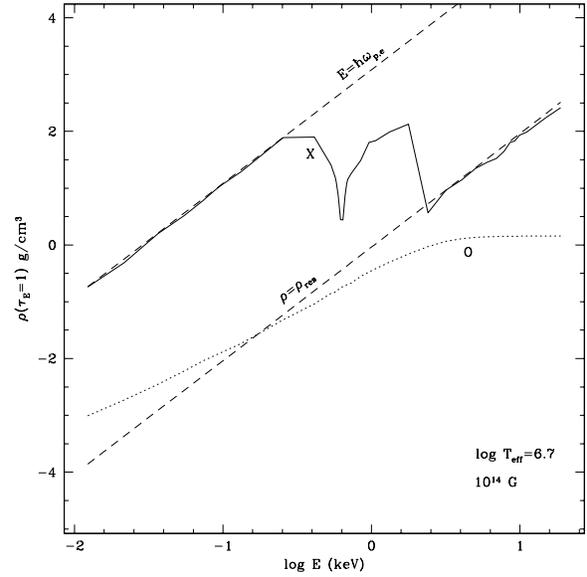}
\caption{ The decoupling density for $B=10^{14}$ G at $\lgTeff=6.7$,
illustrating the transition to photosphere formation within the vacuum
resonance layer above 1.5 keV.  The X-mode flux below 0.25 keV is
dominated by collective plasma effects.  }
\label{fig:magnetar_14_unitrho}
\end{figure}

\textit{High density effects ---} Two phenomena related to dense
plasma are not accounted for in the opacity formulation of
\S\ref{Sec:Opacity}.  Collective plasma effects are encountered where
$v_{e}>1$ strongly influence the optical and near UV flux of thermal
models with radio pulsar type fields (Fig.
\ref{fig:radio_12_polz_details}); the effects extend to X-ray energies
for the magnetar models (Fig.  \ref{fig:magnetar_14_unitrho}).  In
this regime, the collective response of the plasma should strongly
attenuate propagating waves, and the assumption of radiative transport
may be incorrect to long wavelength radiation.  This is plain in the
magnetar spectra, but is also relevant at optical wavelengths for very
cool stars (e.g. RXJ 1856.35) and aging RPs like Geminga.  Collective
phenomena may be included by way of a structure factor in the scattering
opacity as a first step toward evaluating corrections
to the equilibrium thermal structure.  Secondly, plasma is tightly
coupled when $\Gamma\equiv \left(4\upi n_{e}/3\right)^{1/3}e^{2}/(kT)
>1$, but the Debye-H\"uckel approximation (\ref{eq:DebyeHuckel}) is
only a perturbative correction.  A more plausible model for $P_{Coul}$
could be implemented, although errors in the coupling occur at the
very deepest model strata of the most strongly magnetized atmospheres
where an improved calculation is likely to correct only the X mode
intensities in the spectral tail.

\textit{Ionic cyclotron resonances ---} Our opacity calculation is
derived from the self-consistent inclusion of ions in the plasma
dielectric (\ref{eq:vacplasdielect}); energy dependencies which scale
with magnetic field enter the opacity through the eigenvalues of the
polarization tensor (\ref{eq:polzeigen}).  Some authors favor a
calculation of these eigenvalues in which electrons and protons are
distinct plasma constituents, with effective eigenvalues
(c.f. \citet{Zane00,HoLai01}):
\begin{equation}
\label{eq:badeigen}
\pi^{2}_{p,\mathit{eff}} \simeq \frac{1}{(1-p u_{e}^{1/2})^{2}} +
\frac{(m_{e}/m_{p})^{2}}{(1+p \frac{m_{e}}{m_{p}}u_{e}^{1/2})^{2}}
\end{equation}
As discussed in \S\ref{Sec:Cross} the principal source of ion
cyclotron absorption resides in resonant behavior of the polarization
vector itself and, if the vectors $e_{p}^{j}$ are computed
consistently as described in \S\ref{Sec:Opacity}, the discrepancy of
no particular consequence for radiation $\omega>\omega_{c,p}$; the
opacity to longer wavelength radiation however acquires a scaling
factor $\propto u_{e}^{-2}$ from eqn (\ref{eq:polzeigen}), which is
steeper than the $\sim u_{e}^{-1}$ scaling of (\ref{eq:badeigen}).
Results calculated in the approximation (\ref{eq:badeigen}) are
potentially unreliable at X-ray energies for $B\ga10^{14}\
\mathrm{G}$, and the discrepancy is compounded for admixtures of
helium.

\textit{Warm plasma and other opacity effects ---} The opacity model
of \S\ref{Sec:Opacity} is derived assuming ``cold'' electron plasma
plus vacuum and neglects plasma dispersion which is of interest near
the electron cyclotron line \citep{KirkMesz80}.  We have also
neglected Comptonization which affects the amplitude of the high
energy spectral cutoff \citep{KPS83}.  For the NS model temperatures
discussed herein, a fully relativistic treatment of the plasma
response is unnecessary \citep{Meszbook}.  Calculation of the magnetic
Gaunt factors at transitional field strengths $\ell\sim 1$ are
incomplete as described in \S\ref{Sec:Opacity} and could be improved
by using the calculation of \citet{PavlovPanov76}.

%
%
%
%

%
%

%
%

\end{document}